\documentclass{article}

\usepackage{arxiv}
\usepackage{xcolor}
\usepackage[utf8]{inputenc} 
\usepackage[T1]{fontenc}    
\usepackage{hyperref}       
\usepackage{url}            
\usepackage{booktabs}       
\usepackage{amsfonts}       
\usepackage{nicefrac}       
\usepackage{microtype}      
\usepackage{amsmath}
\usepackage{cleveref}       
\usepackage{lipsum}         
\usepackage{graphicx}
\usepackage{natbib}
\usepackage{doi}
\usepackage{placeins}
\usepackage{caption}

\title{\emph{SimPol}: Simulating polarisation in 
\\political belief networks in European countries}

\date{June 26, 2026}

\usepackage{authblk}

\newbox{\orcid}\sbox{\orcid}{\includegraphics[scale=0.06]{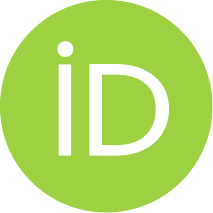}} 

\author[1]{%
	Isabela ~Burattini Freire\thanks{\texttt{ib2920@nyu.edu}}%
}
\author[2]{%
	{\usebox{\orcid}\hspace{1mm}Hongryol ~Cha\thanks{\texttt{h.cha@qub.ac.uk}}}%
}
\author[3]{%
	Irina ~Epure\thanks{\texttt{i.m.epure@liacs.leidenuniv.nl}}%
}
\author[4]{%
	\usebox{\orcid}\hspace{1mm}Sara 
    ~Filippini\thanks{\texttt{sara.filippini@polito.it}}%
}
\author[5]{%
{\usebox{\orcid}\hspace{1mm}Karan~K. H.~Manjunatha}%
\thanks{\texttt{karankhm9@gmail.com}}%
}
\author[6]{%
	Chethan Kavaraganahalli Prasanna\thanks{\texttt{kavaraganahallipra.c@northeastern.edu}}%
}
\author[7]{%
    \href{https://orcid.org/0000-0003-3542-7744}{\usebox{\orcid}\hspace{1mm}
	Ivan
    ~Samoylenko, \thanks{\texttt{uvan63@gmail.com}}}%
}
\author[8]{%
{\usebox{\orcid}\hspace{1mm}~Niels~Van Santen, \thanks{\texttt{niels.vansanten@ugent.be}}}%
}
\author[9]{%
	Adarsh Prabhakaran\thanks{\texttt{adarsh.prabhakaran@rhul.ac.uk}}%
}
\author[10]{%
	Guillermo Romero Moreno
}

\affil[1]{Leipzig University, Germany}
\affil[2]{Queen's University Belfast, UK}
\affil[3]{Leiden University, The Netherlands}

\hypersetup{
pdftitle={A template for the arxiv style},
pdfsubject={q-bio.NC, q-bio.QM},
pdfauthor={David S.~Hippocampus, Elias D.~Striatum},
pdfkeywords={group polarisation, network inference, belief networks, agent-based modelling},
}

\begin{document}
\maketitle

\begin{abstract}

Polarisation is reshaping the political arena across Europe, yet the mechanisms driving it remain difficult to predict, given the complex interplay between individual belief formation, social dynamics, and cultural change.  Here we combine empirical network analysis with agent-based modelling to understand how different ways of structuring belief systems may affect the polarisation drive, and how the diversity of belief systems in Europe may result in different polarisation trajectories. Using the 2016 European Social Survey, we infer belief networks across 23 European countries via a Bayesian algorithm, revealing that belief systems are predominantly organised around immigration, LGBT rights, and economic interventionism, reflecting the influence of populist discourse across the continent. We further verify a Western–Eastern divide across the national belief networks: in Western European countries, left–right self-identification is a more reliable predictor of broader belief alignment, whereas in Eastern Europe this relationship breaks down. By applying these empirical belief networks into a sociologically grounded agent-based model, we further show that polarisation is amplified by high individual belief rigidity and low susceptibility to social influence, and that cross-country differences in polarisation levels mirror the same geographic divide observed in belief network topology. These findings establish belief networks topologies as a structural driver of political polarisation, with implications for understanding and anticipating polarisation dynamics across diverse European contexts.

    We find that populations are not polarised when little attention is placed on maintaining internal coherence and polarisation levels are moderate when high attention is placed in both keeping internal coherence and agreement in beliefs with others.
    When attention to internal coherence is high and social influence is low, a remarkable difference between countries emerges: Western European countries polarise very highly, while Eastern European countries experience more moderate levels of polarisation, an effect that we link to Western European countries having more aligned belief networks across a single dimension.
    Our results point to possible explanations of how the structure of the belief system may play a role in polarising dynamics and how these materialise in Europe.
    This may have implications for policymakers in advancing interventions to reduce social conflicts and potential risks.

\end{abstract}

\keywords{Agent-based modelling \and Belief networks \and European social survey \and Network inference \and Political belief systems}
        

\section*{Introduction}

In recent decades, Europe has experienced a rise in support for extreme political positions. The proportion of Europeans voting for far-right candidates in countries' national elections has grown fivefold since 1995, reaching historic highs by 2025 \citep{Rooduijn2024PopuList}. Simultaneously, an emergent trend in lowering national and institutional trust also observed in the continent, decreasing the political trust by 13.4\% since 2020 \citep{eurofound2022trust}. As the established parliament parties find it hard to reach a social consensus that can resolve extreme political movements \citep{TurnerDArt2024}, these dynamics lead to deeper societal and political polarisation phenomena.

In this context, the attempts to understand the driving forces of this political polarisation have recently increased in the social sciences. One line of research focuses on top--down influences of political elites on voters' attitudes, particularly focusing on the role of mass media in shaping political preferences \citep{HOEWE202019}. Another stream of research understands political polarisation as a bottom--up process emerging from individual cognition through coherence--based reasoning \citep{Jost2022}. Individuals' coherence in reasoning about their political beliefs reflects the degree to which they conform their ideas to what they perceive as the group norm, as opposed to how resistant they remain to social pressure. This dynamic is reflected in the tendency of political discussions to remain confined to close social circles: in Italy (65\%), France (71\%) and Germany (74\%), friends are the main confidantes for political conversations \citep{Ilioaia2025}. Such social confinement reinforces existing views and limits exposure to opposing political beliefs, and might strengthen societal polarisation.

Toward the bottom--up approach to political attitudes, scholars started to view the belief systems of how people think. Belief systems were first conceptualised as structured configurations of attitudes rather than a certain opinion. \cite{Converse1964} conceptualised belief systems as highly structured correlations between political beliefs based on ideologies. In this view, belief systems rely on each other, and cognitive coherence exists when positions on different issues are coherent \citep{fishman2022change,Brandt2021}. For example, people may change political attitudes on one issue and consistently on others based on ideologies. Building on this stream of research, scholars reframe belief systems as the networks of interconnected beliefs \citep{Dalege2017}. In network models, belief systems of attitudes are represented as nodes and their associative relations as edges, which measure coherence through the degree of network stability. 
Central to understanding how individuals navigate social pressure and maintain coherence in their political views is the structure of their belief systems. Considering societies as complex systems \cite{Bliuc2024} 
in this framework, coherence is not a fixed property of an ideology but an emergent feature of the network: individuals tend to reduce the tension arising from dissonant beliefs (pairs of beliefs that, when held together, feel mutually inconsistent) adjusting to the ones that feel more internally compatible or compatible to their social group \citep{Gawronski2012}.

To understand the different pathways by which individuals update their beliefs, respond to social pressure, and ultimately polarise in society, researchers have employed models grounded in both theory and data. Theoretical frameworks in psychology and the social sciences aim to establish generalisable rules that capture how individual characteristics interact with social and physical environments to shape human behaviour \citep{Gilbert2019}. Statistical models look for patterns in data that has already been collected, but can only tell us what has happened in the past, not what might happen under different conditions \citep{ElsenbroichBadham2023}. Agent--Based Models (ABMs) occupy a distinctive middle ground: grounded on data--driven populations' belief networks, combined with testable parameters on how individuals interact with peers and respond to social pressure, they allow researchers to explore theoretical assumptions while keeping the model anchored to real--world evidence \citep{ChattoeBrown2013}. This combination allows ABMs to explore counterfactuals and alternative scenarios --- asking not only what has happened, but what could happen under different conditions --- which is especially valuable when studying polarisation, as the cognitive and cultural forces shaping individual belief updating may themselves shift over time \citep{ElsenbroichBadham2023}. Moreover, by explicitly linking individual--level mechanisms to collective outcomes, ABMs are particularly effective in connecting micro-level assumptions regarding individual agent behaviours to macro-level patterns in societies~\citep{Bliuc2024}.

Prior theoretical work has also emphasised that opinion dynamics should not be reduced to the evolution of a single scalar attitude. Several models represent individuals as holding multiple, interdependent opinions, where belief updating is shaped both by social influence and by internal consistency constraints. \cite{battiston2016interplay}, for instance, proposed a multilayer model in which agents hold opinions on several topics and are simultaneously subject to peer pressure toward local consensus and an internal pressure toward coherence across issues. Related multidimensional extensions of Friedkin--Johnsen and consensus models represent each individual by a vector of opinions, with dependencies between topics encoded by a logic matrix \citep{parsegov2016novel,ye2020continuous}. These models provide important analytical insight into consensus, convergence, and persistent disagreement in systems of interdependent opinions. However, they are typically stylised mathematical frameworks in which the internal structure of beliefs is specified theoretically rather than inferred from empirical belief networks. This leaves open the question of how empirically observed belief--network structures interact with coherence--based reasoning and social influence to generate polarisation in real populations.


In this study, we aim to understand how different belief network structures and coherence--based reasoning jointly give rise to political polarisation across Europe. Specifically, we want to understand how polarisation arises in a country--comparative manner in Europe, understanding the cultural and political dissonances within the continent. To do so, we generate belief networks from survey data across 23 European countries, following the broader belief-network tradition in the study of attitudes and belief systems \citep{BoutylineVaisey2017,Dalege2017,Brandt2021}, and estimate these networks using three complementary network--inference approaches: correlation--based network construction, regularised partial--correlation estimation via the graphical lasso, and minimum--description--length network reconstruction \citep{Masuda2025,EpskampFried2018,Friedman2008,peixotoNetworkReconstructionMinimum2025,grunwaldMinimumDescriptionLength2007}. These belief networks are derived from the European Social Survey (ESS)~\citep{ESS8e02_3}, a cross--national survey measuring citizens' political attitudes and values across Europe. We then feed these empirically grounded belief networks into an Agent--Based Model (ABM) with strong backing from cognitive science \citep{dalege_networks_2025}, in which agents update their beliefs through social interaction and self--reflection. By varying the levels of cognitive coherence across simulations, we explore how these parameters interact with the structure of national belief networks to produce different polarisation dynamics. This approach builds directly on \cite{Batzke2025}, who showed that a stronger drive for coherence leads to greater polarisation in Germany. The present study extends this framework to all 23 countries within the ESS database, allowing us to examine whether these dynamics generalise across Europe.


\begin{figure}
	\centering
    \includegraphics[width=0.7\linewidth]{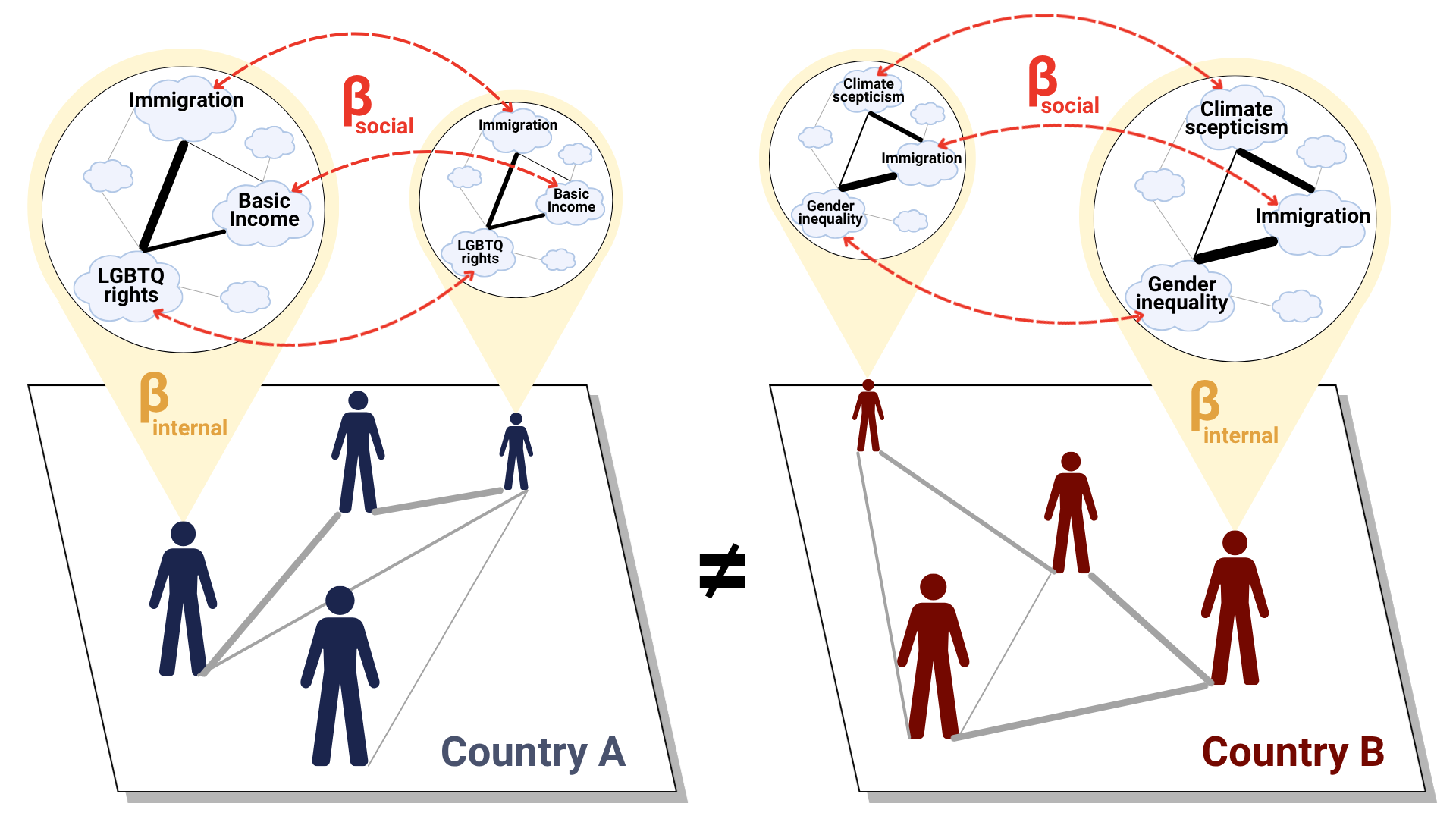}
	\caption{Mental map of methodology.}
	\label{fig:map_pol}
\end{figure}

\section*{Results} \label{sec:results}

\subsection*{Belief networks across countries}

We inferred belief networks for each of the 23 countries in the ESS dataset using Kendall correlations ~\citep{Kendall1938, Masuda2025}, partial correlations ~\citep{EpskampFried2018, Friedman2008}, and nonparametric
Bayesian network reconstruction ~\citep{peixotoNetworkReconstructionMinimum2025} (see Methods section). For further analyses, we focus on results obtained using the Bayesian reconstruction method, which produces sparser networks without dismissing
low weight edges too harshly. More details on this choice are given in Appendix section~\ref{app:compare-inference}. In the resulting networks, nodes represent political beliefs and edges represent the inferred statistical dependencies between them, with their weights reflecting the strength and nature of those correlations - either positive or negative.

\begin{figure}
    \centering

    \begin{minipage}[t]{0.8\linewidth}
        \centering
        \includegraphics[width=\linewidth]{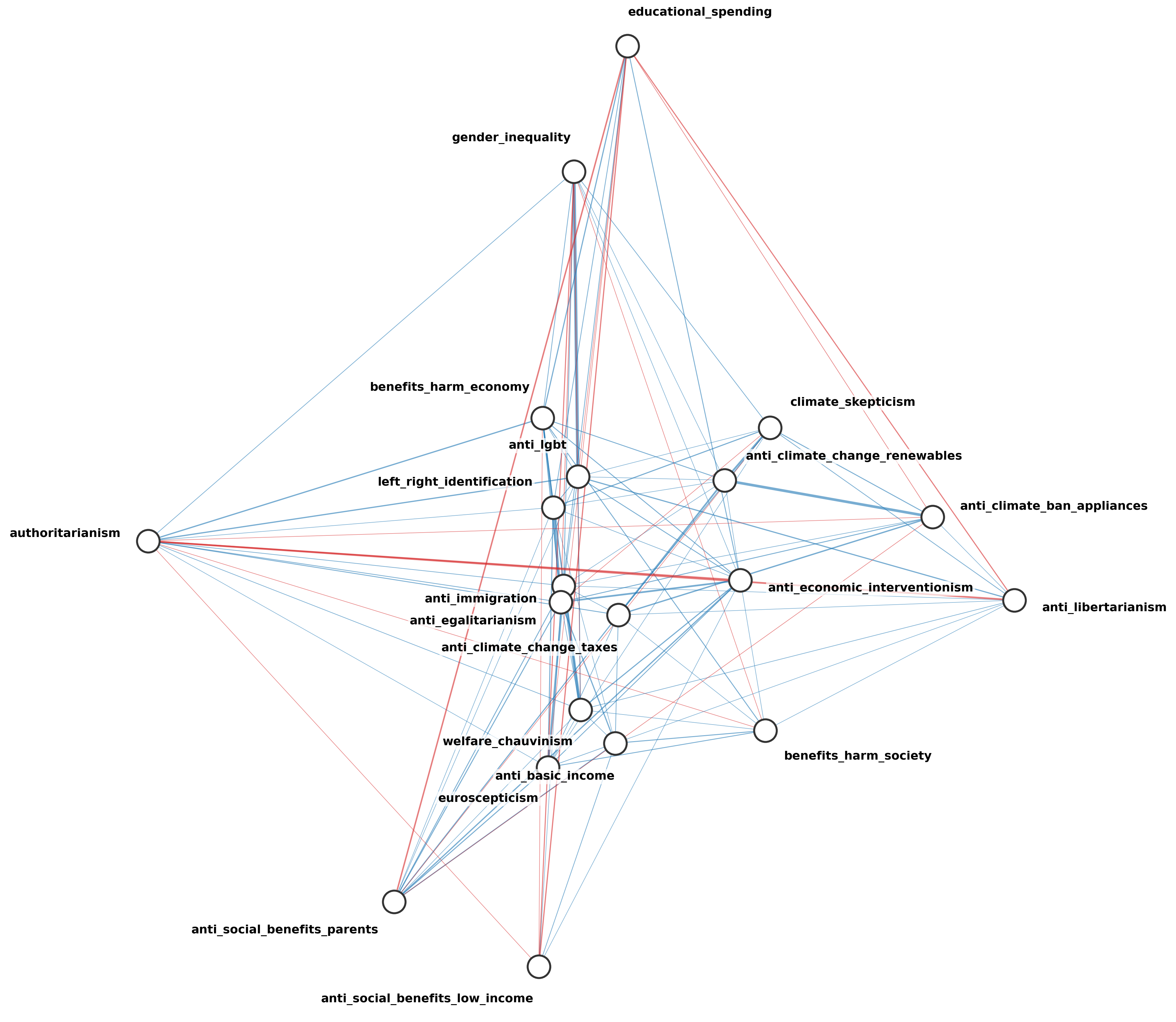}

        \small (a)
    \end{minipage}
    \hfill
    \begin{minipage}[t]{0.8\linewidth}
        \centering
        \includegraphics[width=\linewidth]{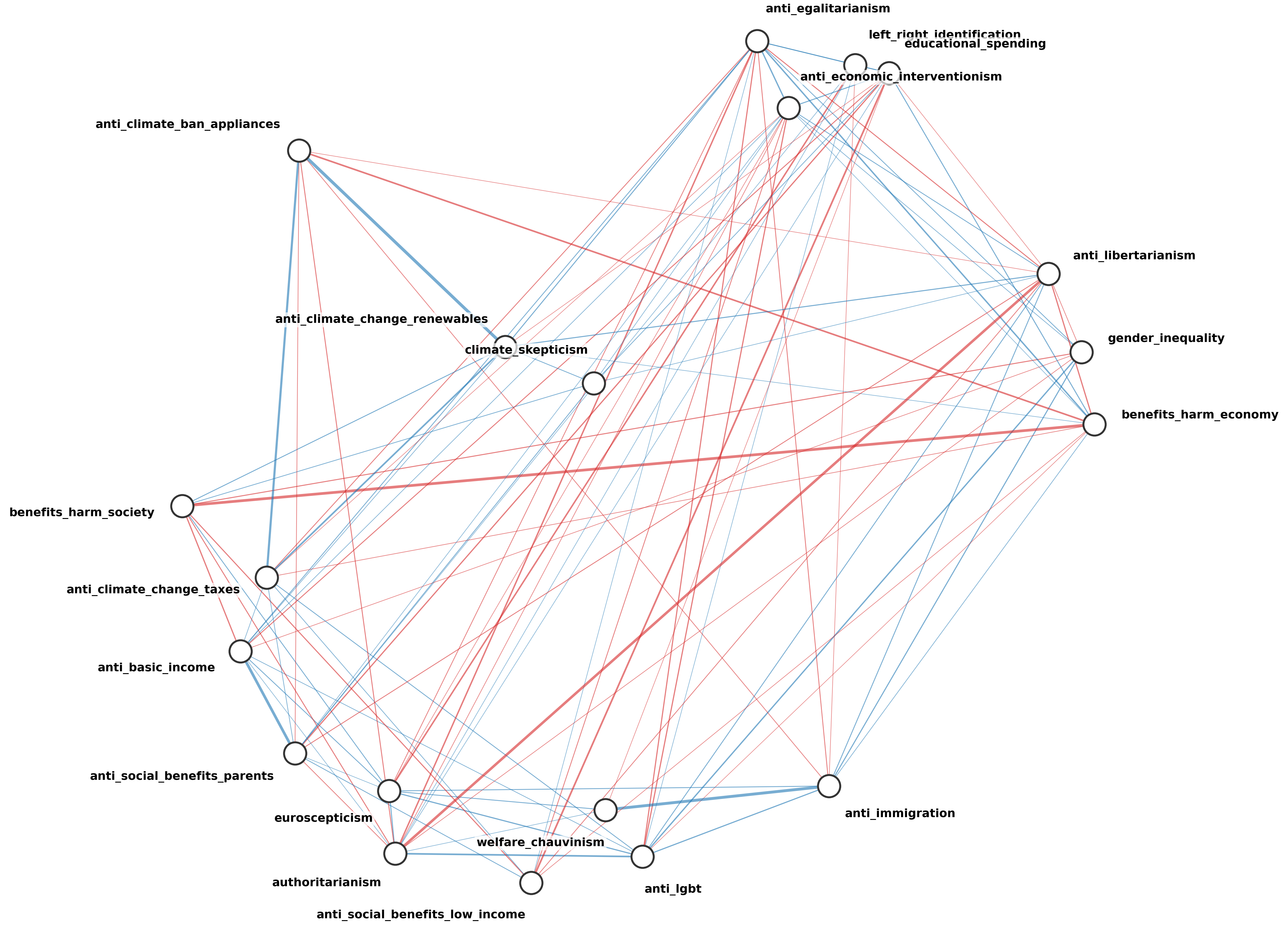}

        \small (b)
    \end{minipage}

    \caption{Comparison between the belief networks of (a) United Kingdom and (b) Russia. Nodes indicate beliefs and edges their correlation. Nodes located further apart denote more negatively correlated beliefs (connected through red edges), while those closer in the network topology are more positively correlated (blue edges).}
    \label{fig:gb_vs_ru}
\end{figure}

For each country, we defined the top five most strongly connected beliefs most strongly correlated with other beliefs (i.e., those with the highest absolute weighted degree sum. We have noticed that certain topics consistently appear in these top five lists across a large number of countries. For example, the topic of immigration is present in the top five list of 18  country networks out of the total 23. This is followed by LGBT rights (17 countries), economic interventionism (15 countries), authoritarianism (14 countries), and egalitarianism (12 countries). Despite this shared presence of beliefs, national belief networks vary substantially in their overall topology. Some countries display a tightly clustered structure in which most beliefs are densely interconnected and a few nodes remain peripheral, while others show a more fragmented pattern in which beliefs divide into two loosely connected groups of roughly equal size. Figure \ref{fig:gb_vs_ru} illustrates this contrast through the networks of Great Britain and Russia. We note that this comparison remains at the level of visual interpretation, as formal clustering analyses of network topology are left for future work.

Extending this interpretation to all belief networks across the 23 countries under analysis, we verify a geographic pattern across the belief network topologies. Networks characterised by a dense central cluster of tightly interconnected beliefs with fewer peripheral nodes - as illustrated by the Great Britain example (Figure~\ref{fig:gb_vs_ru}, Panel a) - tend to characterise Western European countries in the dataset. In contrast, networks where beliefs fragment into two groups of roughly equal size with sparser connections between them — as illustrated by Russia (Figure~\ref{fig:gb_vs_ru}, Panel b) - tend to align with Eastern European countries in the dataset. We also created a webapp to visualize correlations between beliefs and the corresponding belief networks~\footnote{The webpage can be found here: \url{https://complexity72-26-simpol.vercel.app/}}.

\begin{figure}
    \centering

    \begin{minipage}[t]{0.75\linewidth}
        \centering
        \includegraphics[width=\linewidth]{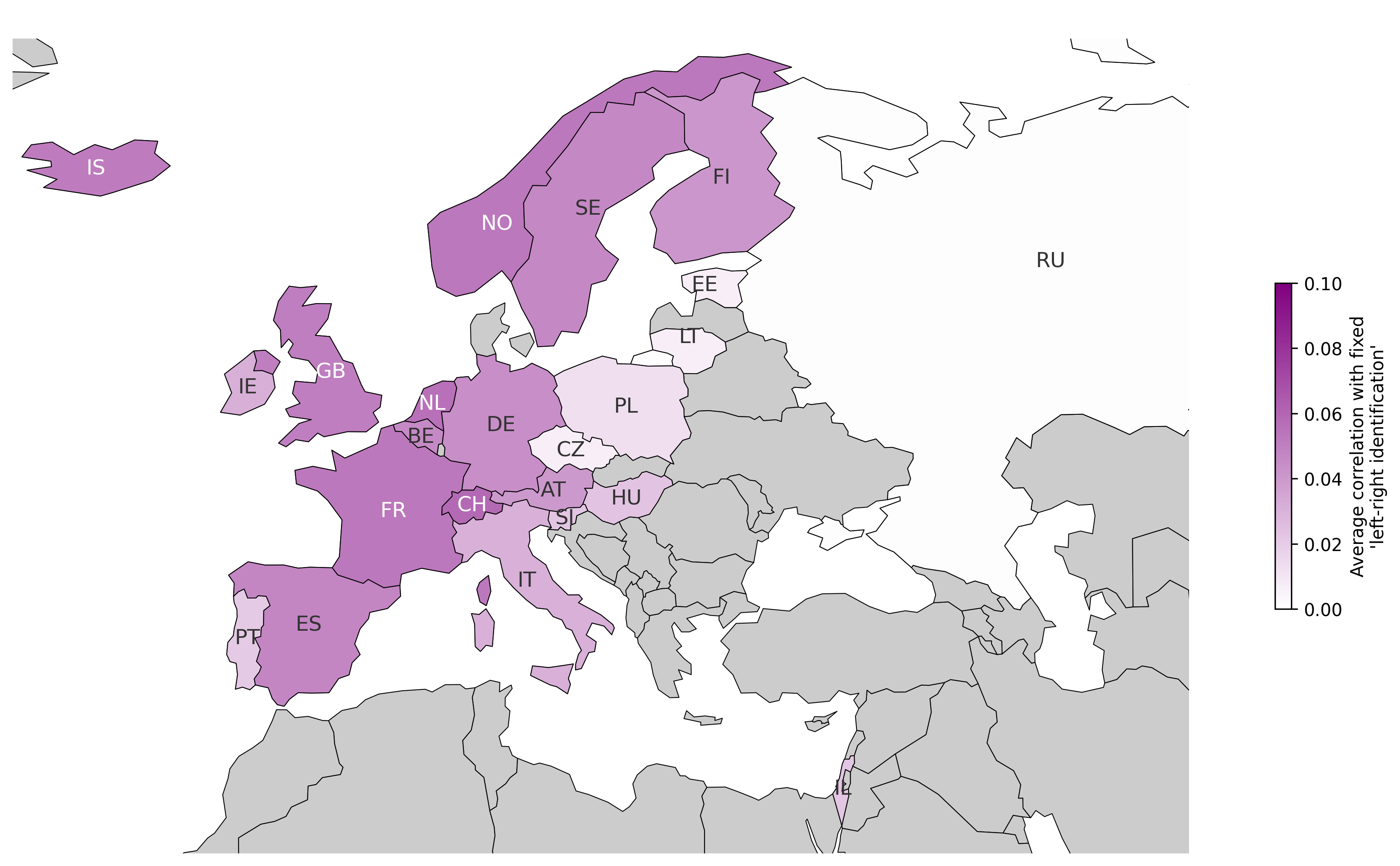}

        \vspace{2mm}
        \small (a)
    \end{minipage}
    \hfill
    \begin{minipage}[t]{0.75\linewidth}
        \centering
        \includegraphics[width=\linewidth]{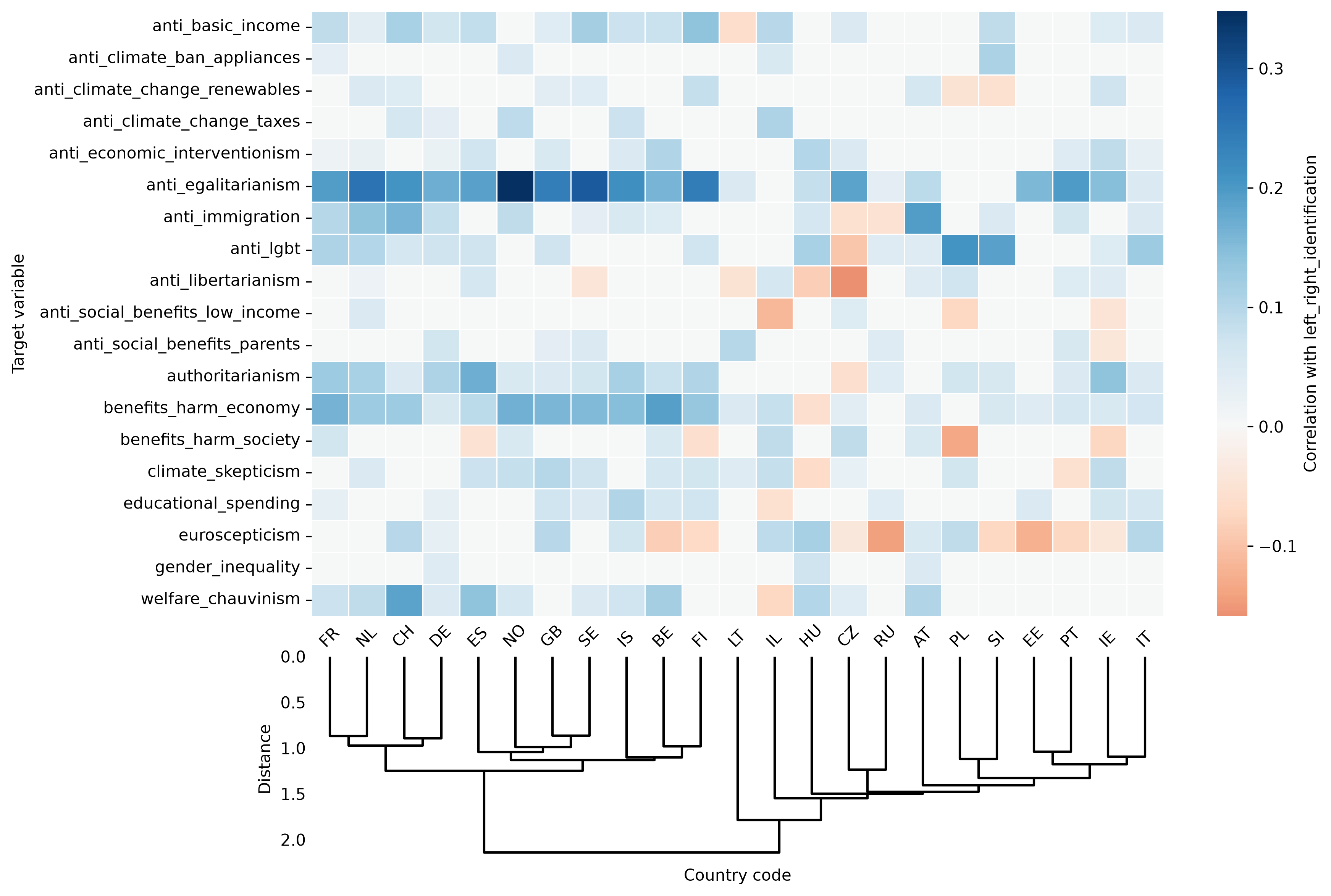}

        \vspace{2mm}
        \small (b)
    \end{minipage}

    \caption{Country-level differences in the association between belief-network structure and left--right self-identification reveal a clear East--West pattern across Europe. (a) Map of the ESS respondent countries showing the average correlation between the nodes in each country's belief network and the left--right self-identification node. Average correlations range from approximately 0.0 to 0.1. (b) Correlations between beliefs derived from the ESS dataset~\citep{VanNoord2025} and the left--right self-identification node. Countries are ordered according to a hierarchical clustering based on the pairwise Frobenius distances~\citep{golubMatrixComputations2013} between their belief networks. The dendrogram displays the resulting clustering. Together, the two plots reveal a near-complete separation between Eastern and Western European countries.}
    \label{fig:map_and_dendogram}
\end{figure}

To further investigate this apparent East--West divide, we specifically examine how the left--right self-identification node correlates with all other beliefs in a pairwise analysis. This allows us to assess how consistently the political spectrum structures the remaining beliefs in each country. For example, if the left--right scale is meaningful in a given country, respondents who place themselves further to the right should tend to differ systematically from respondents who place themselves further to the left across several other belief items. In Western European countries, left--right identification correlates with other beliefs in a consistent direction, suggesting that knowing where someone places themselves on the political spectrum is a potential predictor of their other political beliefs. In Eastern European countries, however, left--right identification correlates with other beliefs in a less consistent pattern, suggesting that the traditional left--right spectrum organises political beliefs less clearly (Figure~\ref{fig:map_and_dendogram}).

\subsection*{Analysis of agent--based simulations}

In the ABM, the behaviour of each individual is controlled by two parameters, $\beta_{pers}$ and $\beta_{soc}$. These regulate attention to the coherence of the personal belief network and to agreement with the social contacts, respectively. Tracking the updated beliefs we use the results of the ABM to measure polarisation in the population. We measured both the per-belief polarisation, using variance to quantify the spread of opinions (appendix~\ref{app:belief_var}) and a more global multi-dimensional polarisation as measured by the variance of the Euclidean distance in belief space (Eq.~\ref{eq:mdpol}). In what follows we focus on the latter measure.

\begin{figure}
	\centering
    \includegraphics[width=0.9\linewidth]{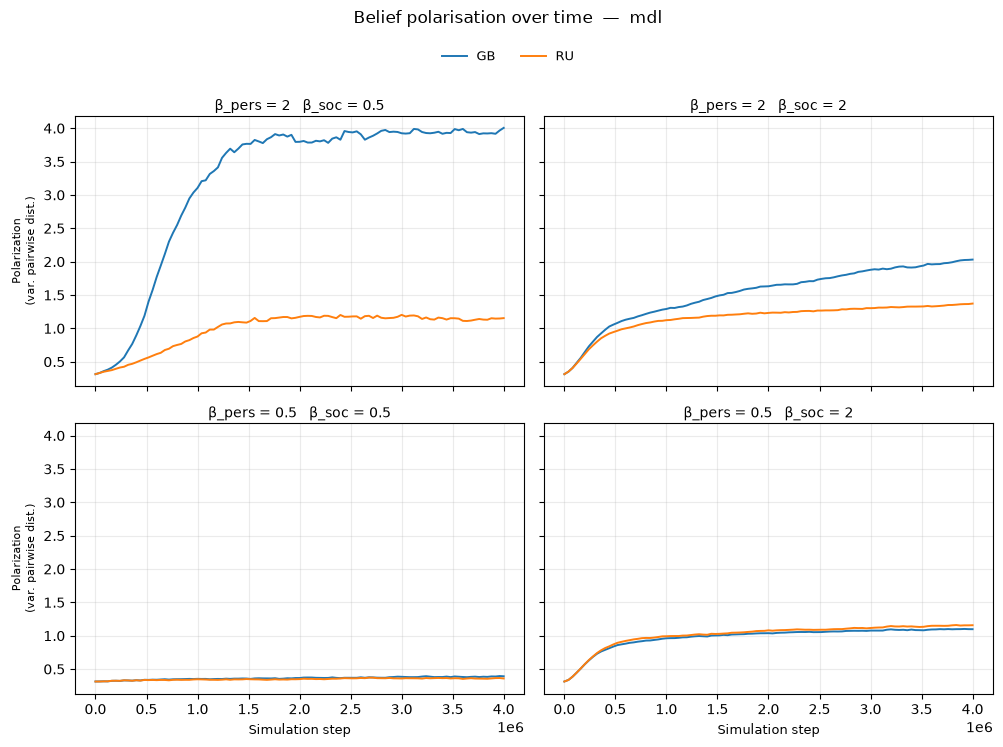}
	\caption{Evolution over time of the polarisation metric for Great Britain (blue line) and Russia (orange line), for four combinations of the attention parameters ($\beta_{pers}$ and $\beta_{soc}$). Each simulation included 2000 identical agents, and was initialized with a random belief vector for each agent. We chose the duration of the simulation such as to allow each agent and each belief to be sampled on average 100 times.}
	\label{fig:pol_traj}
\end{figure}

We ran the ABM for all countries and for all combinations of attention parameters $\beta_{pers} = \beta_{soc} \in \{0.5,2\}$. For the belief network generated by minimum description length, Figure~\ref{fig:pol_traj} shows how the polarisation changes with time in the case of Great Britain and Russia. From the bottom row of Figure~\ref{fig:pol_traj} we can see that when $\beta_{pers}$ is low, both countries develop similar and low levels of polarisation, irrespective of the attention of the agents to each other's opinions. On the other hand, higher values of attention to internal dissonance separate the behaviour of Great Britain and Russia. It generally leads to an increase in polarisation with Great Britain reaching higher values of polarisation for any value of $\beta_{soc}$. In the high $\beta_{pers}$ regime, Great Britain and Russia differ in how their polarisation changes with respect to the attention to social dissonance. For Great Britain an increase in $\beta_{soc}$ at high values of $\beta_{pers}$ significantly reduces the value of the measured polarisation, while Russia sees a slight increase.

Figure~\ref{fig:pol_traj} suggests that differences between countries are more pronounced in the high $\beta_{pers}$ and low $\beta_{soc}$ region, where the topology of the internal belief network has the strongest impact on the dynamics. In Figure~\ref{fig:map_pol}, we show the polarisation value reached at the end of a simulation with $\beta_{pers}=2$ and $\beta_{soc}=0.5$, for each country on a map of Europe. We can clearly note a West-East divide where the Eastern European countries show meaningfully lower values of polarisation.
\begin{figure}
	\centering
    \includegraphics[width=0.7\linewidth]{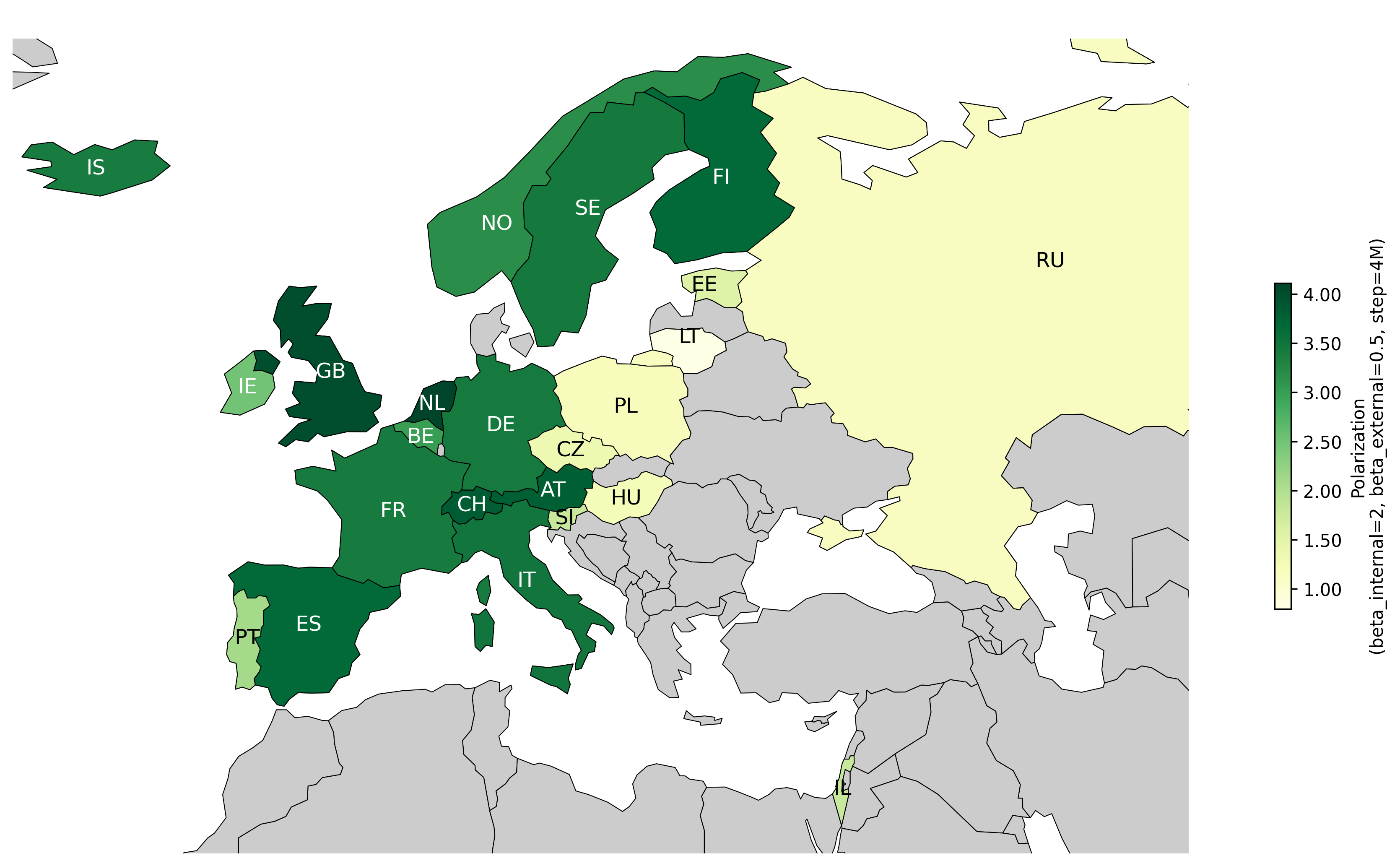}
	\caption{Values of the polarisation metric at the end of the simulations with high $\beta_{pers}$ and low $\beta_{soc}$, for different European countries. Each simulation included 2000 identical agents, and was initialized with a random belief vector for each agent. We chose the duration of the simulation such as to allow each agent and each belief to be sampled on average 100 times.}
	\label{fig:map_pol}
\end{figure}

\begin{figure}
	\centering
    \includegraphics[width=0.6\linewidth]{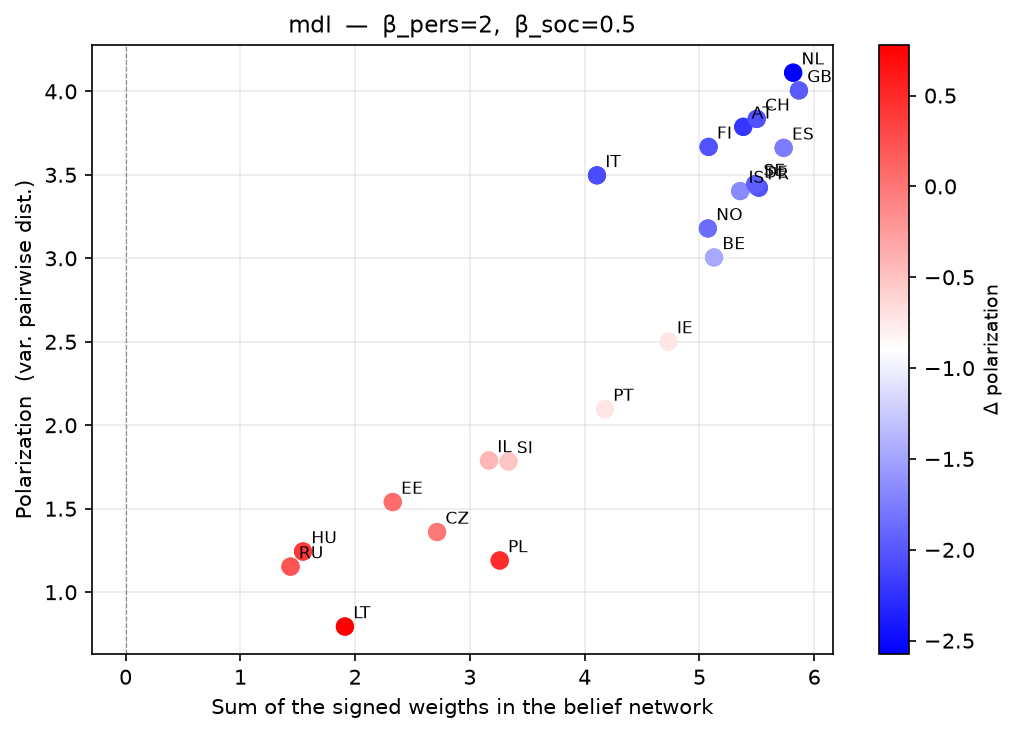}
	\caption{Scatter plot showing the relation between the final polarisation value reached at high $\beta_{pers}$ and low $\beta_{soc}$, and the sum of all signed weights in the country's belief network. The colour scale represents the change in polarisation arising from an increase in $\beta_{soc}$. Each simulation included 2000 identical agents, and was initialized with a random belief vector for each agent. We chose the duration of the simulation such as to allow each agent and each belief to be sampled on average 100 times.}
	\label{fig:scatter_corr_pol}
\end{figure}

Considering the observations in the previous section concerning the different structures of the belief networks across the ESS, we examine whether it is possible to link the polarisation behaviour to some belief network metrics. In Figure~\ref{fig:scatter_corr_pol}, we relate the sum of the edge weights of the internal belief networks of each country with its polarisation at high $\beta_{pers}$ and low $\beta_{soc}$. We note a positive trend where those Eastern European countries with lower polarisation also have internal belief networks that are characterised by smaller sums of the strength of their network nodes. The colour scale shows the impact of increasing $\beta_{soc}$ from 0.5 to 2 on the polarisation of the system. We see that the two groups tend to collapse into a single state of moderate polarisation, as shown also in the upper-right panel of Figure~\ref{fig:pol_traj}.

\section*{Discussion} \label{sec:discussion}
We have arrived at several findings that help us to understand better the relation between how country populations structure their beliefs and their polarisating drive.
First, we have applied a network inference method that produces belief networks with fewer confounder effects and analysed their differences across European countries from the ESS survey.
In our analysis, we discovered stark differences between belief networks, aligning broadly with Western--Eastern European blocks as a common separation in the European political space stemming from the context of the former Cold War era \citep{wolff_inventing_1994}. Our findings materialise more specifically in a higher alignment with policy issues into a single left--right dimension in Western countries, whereas Eastern countries show weaker alignment of beliefs into this single ideological dimension. We consider that this difference may be related to the different perceptions and affections of ideologues in dividing the left--right placement. Interestingly, some of the beliefs that appear as more central in these belief networks are related to some of the populist narratives (e.g., immigration, LGBTI+ rights) that have emerged in Europe during 2016, pointing out that beliefs in these topics were central in defining political dimensions \citep{norris_cultural_2019, mudde_populism_2017}.

Next, we have explored the effect of the differences of these belief networks on polarisation dynamics by running a cognitive--based agent--based model that combines the strive to increase internal coherence among an individual's beliefs and the effects of social pressures \citep{dalege_networks_2025}.
As the model includes parameters to tune the level of attention to each of these two sources of dissonance ($\beta_{pers}$ and $\beta_{soc}$, respectively), we explore possible combinations of these. Expectedly, in the baseline conditions (low attention to both internal coherence and social pressure)
no polarisation emerges and the system settles into a state similar to the initial conditions. Polarisation may then emerge through two separate mechanisms. At low $\beta_{pers}$ and high $\beta_{soc}$
, moderate polarisation is observed, likely following the clustering of the social networks and simulating the creation of in--groups  \citep{mutz_consequences_2002}. Conversely, at high $\beta_{pers}$ and low $\beta_{soc}$
, polarisation appears as a result of the focus on internal coherence, which tends to push individuals towards more extreme positions, as observed in previous studies \citep{dalege_networks_2025, dalege_toward_2015}.

The impact of this second mechanisms, however, is highly dependent on the structure of the belief network and following the previously described East--West divide. While Western countries showing extreme values of polarisation in this parameter range, in Eastern countries the structure of the belief system does not have such a strong polarising drive. In the high $\beta_{soc}$, high $\beta_{pers}$ range,
the polarisation driven by internal coherence is partly countered by a higher interest in mutual agreement, leading to a reduction of polarisation in Western countries, while polarisation remain roughly the same in Eastern countries. 

These results hint to some hypotheses on mechanisms that may be behind some of the polarising dynamics that have been observed in recent years.
The structure of belief systems may have a significant effect on the polarising drive. In our model, three conditions are required to obtain the maximum values of polarisation. A belief structure reflecting a population that aligns strongly along a single political axis, a strong wish for coherency with their internal belief network, and low attention others' opinions. These are conditions that have been observed to have increased in some Western countries recently, where populist rhetorics have increased the emotional and political involvement across Europe \citep{grzymala2016east}, social isolation levels have been on the rise \citep{anttila2020disconnected}, and a high alignment across the left--right dimension is present \citep{coman2017dimensions, lachat2018way, freire2013western}. 
On the contrary, these conditions may not lead to polarising dynamics in Eastern European countries, possibly due to a political belief space not fully aligned into a single dimension and being more fractured \citep{coman2017dimensions, freire2013western}. However, polarisation may be resulting in those countries as a result of other mechanisms, such as increases in authoritarian elites \citep{anna2020ethnopopulism, enyedi2020right}.



Our study also theoretically contributes to other strands of previous literature. Since the seminal work, research on political belief systems has evolved from a focus on ideological constraint and coherence to a more relational and multidimensional understanding of how people organise political attitudes. Literature has evolved through three streams of research on ideological belief systems, cultural-economic belief systems, and unstructured belief systems. First, the structured ideological constraint view is foundational with a one--dimensional focus on left--right ideological structure \citep{Converse1964, fishman2022change}. Second, the two--dimensional approach of cultural and economic systems highlighted the correlations of specific political agenda in more dimensions, mostly divided by cultural and economic issues \citep{Baldassarri2014}. However, subsequent works on cognitive coherence \citep{Batzke2025} and tightness/consensus \citep{Bertero2025} focused on structural inconsistency in political belief systems, rather than the dimensional approach. This correlation--based approach made the third stream advance the unstructured belief systems to address a theoretical anomaly between cognitive coherence and incoherence \citep{VanNoord2025}. While traditional dimensional approaches treat the mismatches between attitudes as simply non-structure or non-attitudes, more recent research has started to extend the idea that apparent misalignment in a conventional ideological dimension may nonetheless reflect alternative internal combinational logics of cognitive coherence \citep{Groenendyk2022}. Our study contributes to this line of research stream on political polarisation from the perspective of the networks of belief systems. Although \cite{Batzke2025} have used the notions of belief systems to address polarisation, their work focuses on how heterogeneity of belief structures within a single population may affect polarising dynamics, using a narrow set of politicdal issues.

Subsequent applications to the networks of political belief systems have shown that ideological belief systems can be understood as emergent from such network structures \citep{dalege_networks_2025}. Networks of belief systems theory is based on three essential premises. First, a single belief depends on other related beliefs as an internal belief network \citep{Brandt2021}. One person’s belief can also depend on the others’ beliefs as an external belief system. Second, people tend to reduce dissonance from the emergent inconsistency among any internal and external beliefs \citep{Gawronski2012}. Third, the amount of attention to cognitive inconsistencies can moderate the potential dissonance \citep{Kitayama2004}. By linking these internal dynamics to external networks of actual others’ beliefs and perceived social beliefs, the theory of the networks of beliefs shifts the focus of belief system research from fitting individuals into structured ideological dimensions to understanding how different network topologies and dissonance reduction behaviours affect coherence and incoherence in political belief systems \citep{dalege_networks_2025}, while this work has not applied the networks of belief systems to the polarisation. In line with this research stream, our analysis of polarisation with the agent-based model advances the theory that the networks of belief systems can be reinforced between internal belief networks and external belief networks in this context. 

Although our findings show a consistent explanation of political polarisation through the lens of the networks of belief systems, it might have some limitations in empirical settings. One limitation exists in the dataset that we used, the 2016 ESS survey data with a cross-sectional dataset. While this dataset is limited to capturing the European belief systems across 23 countries in 2016, it would still be worthwhile to investigate whether this politically polarising outcome would be found across more countries at different times if similar questions were asked in the next round of the ESS survey. While the ESS survey strives to obtain a representative sample of the country population, biases still exist in the self--selection process of who was contacted and who responded to the survey once contacted. While the dataset includes post-stratification weights computed by the ESS team based on gender, age, education, and region (where available), their inclusion in the network inference methods is not straightforward, adding caveats to the generalisation of the results to the country level. Thus, it would be valuable for future research to devise ways to include this sort of corrections into the belief network inference methodology.

Another limitation might be in the network inference methodology. We are inferring the same belief network for the whole population, while this is an unrealistic assumption \citep{Brandt2021}. In reality, each individual may have their own internal belief network, but one would expect very similar belief networks shared by parts of the population. One possibility is to stratify the data into socio-demographic groups where we would expect differences in the structuring of beliefs (e.g., by age). Alternatively, data--driven clustering approaches can be used to extract sub-groups based on their beliefs \citep{Batzke2025,VanNoord2025}.

There is a final limitation of using ABM. We have assumed agents are homogeneous in their internal and social attentions, while different levels of attention would be expected to co-exist in a society. Even more, the answers to political engagement survey questions from the ESS data might be used as proxies to infer the distribution of attention to internal coherence in each country. However, social attention would likely require other data sources.

We have used a small--world network as the social network. This network has some basic characteristics of real social networks, but more sophisticated network models -- for example, networks with community structure or empirically calibrated degree distributions - could be explored instead. 
Equally, the data has not been used to calibrate the model and beliefs of each agent have been initialised at random from the $[-1,+1]$ interval. It would be interesting to investigate how the snapshot of the beliefs presented in the data in each country also affect the resulting polarisation dynamics. However, this would require a more careful exploration and characterisation of the belief distributions at the start and end of the simulations, and we leave for future work. 

We have investigated the effect of a specific mechanism --- the personal and social network structure of the belief systems --- on political polarisation across 23 European countries. By using complementary methods of network inference modelling and ABM, we hope to contribute to capturing the underlying mechanisms of polarisation from a broader perspective on both person beliefs and their reinforcing dynamics through collective networks of social belief systems. This may give some practical implications for policymakers in managing the social conflicts and welfare in ever-changing European societies. By providing further understanding of these specific mechanisms to the interdisciplinary communities studying political polarisation, our study advances the general understanding of the personal and social beliefs drivers to address political polarisation in the recent drifts in European countries. We hope that this may also help in proposing measures and lines of action that may reduce the potential risks and social costs of political polarisation.







\section*{Methods} \label{sec:methods}

\subsection*{Data and pre-processing}

We used data from Round 8 of the European Social Survey (ESS), integrated file edition 2.3 \citep{ESS8e02_3}. The analysis follows the data structure used by \citep{VanNoord2025}, who examined political belief systems across 23 European countries using Correlational Class Analysis (CCA). The raw ESS Round 8 file contained (44{,}387) respondents. Following the preprocessing strategy described by \citep{VanNoord2025}, we constructed an initial CCA-ready dataset containing (37{,}118) respondents from 23 countries and 20 political belief variables.
The countries included in the final dataset were Austria, Belgium, Switzerland, Czechia, Germany, Estonia, Spain, Finland, France, the United Kingdom, Hungary, Ireland, Israel, Iceland, Italy, Lithuania, the Netherlands, Norway, Poland, Portugal, the Russian Federation, Sweden, and Slovenia.


The final dataset contains 20 political belief variables selected from the ESS Round 8 questionnaire. These variables cover left--right identification, gender inequality, LGBT attitudes, European integration, immigration, egalitarianism, welfare-state attitudes, economic interventionism, climate-policy attitudes, climate scepticism, authoritarianism, and libertarian values. The 20 belief variables are listed in Table~\ref{tab:belief_variables}.
All belief variables were rescaled to the interval ($[0,1]$). The coding direction was chosen so that higher values indicate more right-wing, conservative, or anti-progressive responses, following \citep{VanNoord2025}. For example, higher values indicate stronger anti-immigration attitudes, stronger opposition to LGBT rights, stronger Euroscepticism, stronger opposition to climate-change policies, and stronger authoritarian value orientations. Conversely, lower values indicate more left-wing, progressive, or pro-egalitarian responses.
Some belief variables were based on a single ESS item, whereas others were constructed from multiple ESS items. For multi-item variables, the variable score was calculated as the mean of the corresponding recoded ESS items. Each underlying item was first cleaned, reverse-coded when necessary, and rescaled to the interval ($[0,1]$). Invalid ESS response codes, including refusal, don't know, no answer, and not applicable responses, were coded as missing values.
The dataset also retains standard demographic variables (age, gender, education, income, religious belonging, urbanisation, and voting participation), which are not used in the present analysis- their coding is described in Supplementary~\ref{app:demographics}.

\begin{table}[htbp]
\centering
\caption{Constructed political belief variables used in the initial CCA-ready dataset.}
\label{tab:belief_variables}
\begin{tabular}{p{0.30\textwidth}p{0.20\textwidth}p{0.40\textwidth}}
\hline
\textbf{Belief variable} & \textbf{Underlying ESS item(s)} & \textbf{Higher values indicate} \\
\hline
Left--right identification & \texttt{lrscale} & More right-wing self-placement \\
Gender inequality & \texttt{mnrgtjb} & More gender-inegalitarian attitudes \\
Anti-LGBT & \texttt{freehms}, \texttt{hmsfmlsh}, \texttt{hmsacld} & More anti-LGBT attitudes \\
Euroscepticism & \texttt{euftf} & More Euroscepticism \\
Anti-immigration & \texttt{imsmetn}, \texttt{imdfetn}, \texttt{impcntr} & More anti-immigration attitudes \\
Anti-egalitarianism & \texttt{gincdif}, \texttt{dfincac}, \texttt{smdfslv} & More anti-egalitarian attitudes \\
Benefits harm economy & \texttt{sbstrec}, \texttt{sbbsntx} & Stronger belief that social benefits harm the economy \\
Benefits harm society & \texttt{sbprvpv}, \texttt{sbeqsoc} & Stronger belief that social benefits harm society \\
Welfare chauvinism & \texttt{imsclbn} & More restrictive views on immigrants' access to welfare \\
Anti-economic interventionism & \texttt{gvslvol}, \texttt{gvslvue}, \texttt{gvcldcr} & Less support for government economic intervention \\
Anti-social benefits: low income & \texttt{bnlwinc} & More restrictive views on low-income benefits \\
Anti-social benefits: parents & \texttt{wrkprbf} & More opposition to benefits for working parents \\
Educational spending & \texttt{eduunmp} & More support for education spending tradeoff \\
Anti-basic income & \texttt{basinc} & More opposition to basic income \\
Anti-climate-change taxes & \texttt{inctxff} & More opposition to fossil-fuel taxes \\
Anti-climate-change renewables & \texttt{sbsrnen} & More opposition to renewable-energy subsidies \\
Anti-climate ban of appliances & \texttt{banhhap} & More opposition to banning inefficient appliances \\
Climate scepticism & \texttt{ccnthum} & Stronger climate scepticism \\
Authoritarianism & \texttt{impsafe}, \texttt{ipfrule}, \texttt{ipbhprp}, \texttt{ipstrgv}, \texttt{imptrad} & More authoritarian values \\
Anti-libertarianism & \texttt{impdiff}, \texttt{ipadvnt}, \texttt{ipcrtiv}, \texttt{impfree}, \texttt{ipudrst} & Less libertarian value orientation \\
\hline
\end{tabular}
\end{table}


Following the 
procedure described by \citep{VanNoord2025}, respondents were retained if they had at most two missing values among the 20 belief variables. In other words, each retained respondent had valid responses for at least 18 of the 20 constructed belief variables.
Respondents younger than 18 were excluded. Respondents whose age was missing were retained if they satisfied the belief-variable missingness rule, because age was used only as a demographic variable and not as one of the 20 belief variables. 
The resulting initial 
dataset contained (N=37{,}118) respondents across 23 countries.




To validate the preprocessing, we compared the reproduced descriptive statistics of the 20 constructed belief variables with Supplementary Table A2 of \citep{VanNoord2025}. For each belief variable, we compared the number of non-missing observations, the mean, and the standard deviation. The reproduced non-missing counts matched the reported values, and the reproduced means and standard deviations matched the published values when rounded to two decimal places. 

\subsection*{Network inference from belief associations}
\label{subsec:network_inference}

We inferred belief networks from the preprocessed European Social Survey Round 8 data. Following the belief-system and CCA literature, political beliefs were represented relationally: belief variables were treated as nodes, and statistical associations between pairs of belief variables were treated as edges \citep{Boutyline2017,BoutylineVaisey2017,Barbet2020,Keskinturk2022,VanNoord2025}. This representation is also consistent with the broader network approach to belief systems, where the structure of associations among attitudes is the main object of study~\citep{dalege_networks_2025,TurnerZwinkelsBrandt2022}. At this stage, the inferred networks should therefore be interpreted as empirical association networks, not as causal networks.

Let \(g\) denote a respondent group, such as a country sample, and let
\begin{equation}
\mathbf{X}^{(g)}
=
\left[
x^{(g)}_{ij}
\right]
\in [0,1]^{n_g \times p}
\label{eq:belief_response_matrix}
\end{equation}
be the belief-response matrix for that group. In Eq.~\eqref{eq:belief_response_matrix}, \(n_g\) is the number of respondents in group \(g\), and \(p=20\) is the number of constructed belief variables. Rows correspond to respondents and columns correspond to belief variables. All belief variables were coded on the interval \([0,1]\), with larger values indicating more right-wing, conservative, or anti-progressive responses, following the coding direction used in the ESS belief-system analysis of \citep{VanNoord2025}. Identifier variables, demographic variables, voting variables, and missingness-count variables were not treated as nodes in the belief network.

For the pairwise correlation network, we computed Kendall's rank correlation coefficient between each pair of belief variables~\citep{Kendall1938} and removed self-correlations. Let \(\mathbf{X}^{(g)}_{\cdot j}\) denote the \(j\)-th column of the belief-response matrix \(\mathbf{X}^{(g)}\), that is, the vector containing the responses of all \(n_g\) respondents in group \(g\) to belief variable \(j\):
\begin{equation}
\mathbf{X}^{(g)}_{\cdot j}
=
\left(
x^{(g)}_{1j},
x^{(g)}_{2j},
\ldots,
x^{(g)}_{n_g j}
\right)^{\top}.
\label{eq:belief_column_vector}
\end{equation}
Here, the dot in the subscript \(\cdot j\) indicates that all rows of column \(j\) are used. For belief variables \(j\) and \(k\), the pairwise association was then defined as
\begin{equation}
R^{(g)}_{jk}
=
\operatorname{cor}_{K}
\left(
\mathbf{X}^{(g)}_{\cdot j},
\mathbf{X}^{(g)}_{\cdot k}
\right),
\qquad
j,k=1,\ldots,p.
\label{eq:kendall_pairwise_association}
\end{equation}
In Eq.~\eqref{eq:kendall_pairwise_association}, \(\operatorname{cor}_{K}\) denotes Kendall's rank correlation coefficient. Thus, \(R^{(g)}_{jk}\) measures the rank-based association between belief variables \(j\) and \(k\) across respondents in group \(g\). 


Kendall's rank correlation was used for three reasons. First, the ESS belief variables are constructed from ordinal survey responses and rescaled composite variables, so the numerical distance between adjacent response categories should not be assumed to have the same interpretation for all items. Kendall's correlation is based on the ranking of responses and on the concordance or discordance of respondent pairs; it therefore does not require the same interval-scale interpretation as Pearson correlation. Second, Kendall's correlation measures monotonic association rather than only linear association, which is appropriate when two belief variables may move in the same or opposite direction without following a strictly linear relationship. Third, the rank-based construction reduces sensitivity to marginal scaling choices and extreme values, which is useful because the constructed belief variables are bounded on \([0,1]\). In the current implementation, respondents with missing values in any selected belief variable were removed before computing the Kendall correlation matrix.

The resulting group-specific correlation matrix is
\begin{equation}
\mathbf{R}^{(g)}
=
\left[
R^{(g)}_{jk}
\right]
\in [-1,1]^{p \times p}.
\label{eq:group_correlation_matrix}
\end{equation}
The matrix \(\mathbf{R}^{(g)}\) is symmetric, with \(R^{(g)}_{jj}=1\). Positive entries indicate that respondents who score higher on one belief variable also tend to score higher on the other. Negative entries indicate that higher scores on one belief variable tend to be associated with lower scores on the other.

The corresponding weighted signed belief network was defined as
\begin{equation}
G^{(g)}
=
\left(
V,
E^{(g)},
W^{(g)}
\right),
\label{eq:weighted_signed_belief_network}
\end{equation}
where the node set is $V=\{1,2,\ldots,p\}$.
Here, \(V\) is the set of belief variables, \(E^{(g)}\) is the set of inferred associations among belief variables in group \(g\), and \(W^{(g)}\) is the weighted adjacency matrix.

As an alternative network-inference procedure, we also implemented partial-correlation networks using a Gaussian graphical model estimated by graphical lasso \citep{Friedman2008,EpskampFried2018,Borsboom2021}, which retains an edge between two beliefs only if they remain associated after conditioning on all other beliefs. Missing values were imputed by the median and variables standardised. The full derivation is given in Appendix~\ref{app:lasso}.


The two network-inference procedures answer different questions. The pairwise Kendall correlation network asks whether two beliefs co-vary across respondents in the selected group. The partial-correlation network asks whether two beliefs remain associated after statistically controlling for all other beliefs in the same model.

One way to overcome the challenge of reconstructing networks from correlations is to use a generative model approach. Using the minimum description length principle~\citep{grunwaldMinimumDescriptionLength2007}, a statistically-grounded nonparametric Bayesian network reconstruction model to infer sparse networks from behavioural data is used to infer belief networks from the ESS data~\citep{peixotoNetworkReconstructionMinimum2025}. We infer belief systems of all 23 countries in our dataset, where each belief system is represented as a network. Nodes in the network are individual beliefs derived from the ESS~\citep{VanNoord2025} and the edges are signed and weighted, with the weights being the partial correlations between pairs of beliefs. The partial correlations are computed from the precision matrix inferred in the model, as done earlier in the graphical-lasso model.

For cross-country analyses, the same procedure was repeated separately for each country. For each country-specific subset, our pipeline computed the pairwise Kendall correlation matrix, the graphical-lasso partial correlation matrix, and the Bayesian reconstructed matrix.

\subsection*{Agent--Based Model}

To simulate the interactions between individuals in a society, we constructed an ABM, based on a statistical physics framework~\citep{dalege_networks_2025}. In analogy with a physical spin model, each individual seeks to lower their potential energy arising from disagreement with its neighbours (social dissonance), as well as lack of coherence of the internal belief network (personal dissonance). 

Each agent $k$ is characterised by an array of $m=20$ discrete beliefs denoted by $\mathbf{b}^k$. At each iteration, a randomly selected agent discusses a randomly selected belief $i$, which is then updated to a new available state according to the following transition probability:

\begin{equation}
    P(b^k_i\rightarrow b'^k_i)=\frac{1}{1+e^{\Delta D}}/\sum_m\frac{1}{1+e^{\Delta D}}
\end{equation}

\noindent where $m$ is the number of available states for $b^k_i$. $\Delta D$ represents the change in dissonance experienced by individual $k$ as a result of a transition from $b^k_i$ to $b'^k_i$. Total experienced dissonance $D$ is described as a sum of personal and social experienced dissonances:

\begin{equation}
    D=\beta_{pers}H_{pers}+\beta_{soc}H_{soc}
\end{equation}

\noindent where $\beta_{pers}$ and $\beta_{soc}$ are parameters referring to an individual's attention to the coherence of their own belief system ($H_{pers}$) and the distance between their beliefs and those of their social contacts ($H_{soc}$), respectively. $H_{pers}$ and $H_{soc}$ are defined as follows:

\begin{align}
    H_{pers} &=-\sum_{ij}b^k_ib_j^kw_{ij} \\
    H_{soc} &=-\sum_{l}b^k_ib_i^lA_{kl}
\end{align}

\noindent Here, $w_{ij}$ is the weight of the edge between belief $i$ and $j$ in the internal belief network (see Methods section). $A_{kl}$ is the adjacency matrix with entries equal to 1 if $k$ and $l$ are individuals connected in the social network, and 0 otherwise.

Our ABM was adapted from \citep{dalege_networks_2025}. Compared to the original model, we assumed that each agent has complete knowledge on the beliefs of its social contacts. 
We chose to simplify this model to aid the interpretability of our results, in consideration of the added complexity resulting from the inferred belief networks.


At the beginning of each simulation, we initialise an artificial social network to guide the interactions between agents in the ABM, using the network model proposed in \citep{boccaletti_growing_2007}. This model reproduces the main properties of real-world networks, including a small-world behaviour (for example, high clustering, low average path length), and a power-law tail in the node degree distribution. 


The ABM is run on a network of 2000 agents, all subject to the same attention parameters ($\beta_{pers}$ and $\beta_{soc}$). 
The belief vector of each agent was initialised uniformly at random. 
We ran each simulation for $4 \times 10^6$ iterations, so that each belief of each agent was sampled, on average, approximately 100 times. 
This number of updates was sufficient for the system to stabilise (in terms of all explored measures) in the neighbourhood of some state, while preserving the structural differences induced by the model dynamics. 
Thus, the final configurations can be meaningfully compared in terms of the qualitative structures generated by different parameter settings.


\subsection*{Measuring polarisation}

A key question in opinion dynamics is whether a system exhibits consensus or polarisation. The notion of consensus is relatively intuitive: in general terms, it refers to similarity or proximity between the opinions of different agents, understood with respect to their distribution, multidimensional distribution, variation, or any other relevant characteristics. Polarisation, in contrast, can be understood as a state in which agents' opinions are not similar in some respect. It is also important to specify how opinions are represented. At least three cases can be distinguished. First, opinions may be one-dimensional, usually taking values on an interval such as $[-1,1]$, or multidimensional where the opinion vector may be analysed component-wise. Second, opinions may be treated as genuinely multidimensional objects, so that polarisation is assessed through measures of dissimilarity in the corresponding multidimensional space. Third, the structure of the interaction network may itself be used to describe the presence or absence of polarisation (for both one-dimensional and multi-dimensional case); in this sense, the network structure provides an additional direction of analysis.

In this section, we discuss methodological approaches to identify consensus and polarisation in all three cases described above: one-dimensional opinion distributions, genuinely multi-dimensional opinion representations, and graph-based descriptions of opinion structure.

\subsubsection*{One-dimensional polarisation}

In the one-dimensional case, polarisation is not a single unambiguous property of an opinion distribution. Ref.~\citep{bramson2017understanding} distinguishes nine different senses of polarisation in one-dimensional opinion distributions. These senses capture different aspects of the distribution, such as extremity, dispersion, clustering, separation between groups, agreement within groups, and the relative size of groups. The full list is reported in Appendix~\ref{app:onedim_polarisation}. These senses are not explicitly reducible to one another: an opinion distribution may become more polarised according to one criterion while remaining unchanged, or even becoming less polarised, according to another.

Another related one-dimensional diagnostic is the histogram-based (m)-value, which is used to assess the modality of an opinion distribution \citep{gregg2019frequency}. This measure is useful for detecting whether a one-dimensional distribution is closer to a unimodal or bimodal shape, but it is not directly applicable to genuinely multidimensional opinion representations. In the present model, each agent is described by a vector of interdependent beliefs, and the relevant structure is not only the marginal distribution of each belief, but also the joint configuration of beliefs across the full belief space. Therefore, applying the (m)-value separately to individual belief dimensions would provide only a partial diagnostic and would not capture the multidimensional structure of polarisation generated by the model.

For this reason, we do not use any one-dimensional measure as the main polarisation metric. Instead, one-dimensional measures can be interpreted as component-wise diagnostics that may complement, but not replace, a multidimensional analysis. In the following subsection, we therefore move from one-dimensional measures to polarisation metrics defined on the full belief-vector representation.

The polarisation values shown in Figures~\ref{fig:pol_traj}-~\ref{fig:scatter_corr_pol} are obtained via an intuitive generalisation of considering single belief variances. We calculated the $L_2$-distance between the beliefs of every agent pair $(k,l)$,
\begin{equation} \label{eq:mdpol}
    d_{kl} = \sqrt{\sum_{i=1}^m (b_i^k - b_i^l)^2},
\end{equation}
from which a heuristic measure for polarisation across the entire belief space is given by the variance of $d_{kl}$.

\subsubsection*{Multi-dimensional polarisation}

A natural generalisation of the one-dimensional case is to represent each agent by a multi-dimensional opinion vector. Such representations are well established in the literature on opinion dynamics, especially in models where agents hold positions on several interdependent topics. What is less standardised, however, is the measurement of polarisation as a single scalar quantity in a genuinely multidimensional opinion space.

One example of a genuinely multidimensional polarisation measure is the approach described in Ref.~\citep{gubanov2019multidimensional}. In that model, each agent's opinion is represented as a stochastic vector over several extreme alternatives, and polarisation is measured through a multidimensional index based on the average distance of agents' opinions from the population centre of mass. This approach captures polarisation directly in a multidimensional opinion space. However, it requires the explicit specification of several extreme alternatives, or effectively predefined opinion clusters. Since our model does not assume such predefined alternatives, we do not rely on this type of multidimensional polarisation index in the present work.

Multi-dimensional opinion vectors are also used in other models of opinion dynamics. For example, \cite{parsegov2016novel} represent each agent's opinion as a vector of topic-specific components and introduce the MiDS matrix (C), which encodes interdependencies between issues. Positive couplings in this matrix tend to align topic-specific opinions, whereas negative couplings may lead to their polarisation. Similarly, \cite{baumann2021emergence} consider opinions as vectors in a multi-dimensional topic space and analyse how correlations between different opinion dimensions may emerge.

However, in such works, a general scalar measure of multidimensional polarisation which is essentially different from one-dimensional distributional measures, is usually not introduced. 
Instead, the analysis is typically based on a collection of one-dimensional diagnostics applied to different coordinates of the opinion vector, together with pairwise relations between these coordinates, such as correlations or topic-overlap parameters.

The polarisation values shown in Figures~\ref{fig:pol_traj}-~\ref{fig:scatter_corr_pol} are obtained via an intuitive generalisation of considering single belief variances. We calculated the $L_2$-distance between the beliefs of every agent pair $(k,l)$,
\begin{equation} \label{eq:mdpol}
    d_{kl} = \sqrt{\sum_{i=1}^m (b_i^k - b_i^l)^2},
\end{equation}
from which a heuristic measure for polarisation across the entire belief space is given by the variance of $d_{kl}$.

In addition to the analysis of opinion vectors, one can also study the global structure of opinion matrices. Given an opinion matrix $(X \in \mathbb{R}^{n \times m})$, where rows correspond to agents and columns correspond to issues, a correlation matrix between opinion dimensions can be constructed and its structural properties analysed. In this setting, polarisation can be described indirectly in terms of the dimensional structure of the correlation matrix. For example, a possible signature of ideological polarisation is a decrease in the effective dimensionality of opinions: if many issue or belief positions become aligned with a small number of latent ideological axes, the opinion space effectively collapses to a lower-dimensional structure. This idea is close to the notion of issue alignment discussed in Ref.~\citep{schweighofer2024raising}, where it is argued that polarisation requires the alignment of issue positions into a common ideological spectrum, allowing a high-dimensional space of issues to be reduced to a small number of ideological dimensions.

A related approach is used by Ref.~\citep{vendeville2026political}, who analyse the latent dimensionality of political attitudes. In our case, the inferred opinion matrices are correlation-type matrices, including partial-correlation matrices and related inferred partial-correlation structures as described in Section~\ref{app:compare-inference}. app:compare-inferenceSince these matrices are symmetric but not necessarily positive semidefinite, a standard PCA-like interpretation of the positive eigenvalue spectrum may be inappropriate. We therefore use the effective rank based on the absolute eigenvalue spectrum as a robustness check, with the full definition and procedure reported in Appendix~\ref{app:effective-rank}.

The results are consistent with the main findings of the paper: matrices corresponding to conditionally Western'' countries tend to have lower effective rank than those corresponding to conditionally Eastern'' countries, indicating a more compressed latent opinion structure. Since this analysis is not central to the main methodological contribution of the paper and requires a separate discussion, we report it in Appendix~\ref{app:effective-rank}.

\subsubsection*{Social Graph-based measures}

A further class of approaches measures polarisation through the structure of a social graph, whenever such a graph is available. Here, the nodes of the graph are individuals or agents, and edges represent social ties or interaction channels between them. These methods therefore differ from belief-network measures, where nodes correspond to beliefs and edges represent statistical or cognitive relations between beliefs.

The general idea of social-graph-based measures is to compare the structure of opinions with the structure of social ties. First, agents are assigned to opinion groups, for example by clustering their opinions or by using predefined labels. One then asks whether social ties are aligned with this grouping: polarised systems are expected to have many ties within opinion groups and relatively few ties between them~\citep{phillips2023high}. Common measures in this family include the within/between index, the internal/external index, the E-I index, modularity, and spectral segregation measures.

We do not use these graph-based measures as part of the main analysis. In our setting, the belief data are empirical, but the social graph is not observed. The network of social ties used in the ABM is synthetically generated only to define possible interactions between agents. Applying social-graph-based polarisation measures would therefore introduce an additional modelling layer and would partly evaluate assumptions about the generated social network rather than directly measuring polarisation in the empirical belief space. For this reason, we use belief-space dispersion as the main polarisation metric and treat social-graph-based measures as less applicable to the present empirical setup.

\section*{Acknowledgements}
This work is the output of the workshop Complexity72h by Complexity Next Gen, held at Northeastern University London, London, UK, 22-26 June 2026. \url{www.complexitynextgen.org/complexity72h/}. K.K.H.M. acknowledges support via
a PhD scholarship from the Scuola Superiore Meridionale. A.P. acknowledges support from a Leverhulme Trust International Professorship Grant (LIP-2022-001).

\bibliographystyle{unsrtnat}
\bibliography{references}  






\appendix
\section*{Appendix}

\section{Demographic variables retained in the dataset}
\label{app:demographics}

In addition to the 20 belief variables, the dataset preserves respondent identifiers, country information, and selected demographic variables. These variables are retained for later descriptive or explanatory analyses but are not used to construct the initial belief-variable matrix.

The demographic variables are coded as follows. The variable \texttt{agea} records respondent age in years. The variable \texttt{gndr} is coded as (1=) male and (2=) female. The variable \texttt{eisced} records the original seven-level ESS education scale: (1=) less than lower secondary, (2=) lower secondary, (3=) lower-tier upper secondary, (4=) upper-tier upper secondary, (5=) advanced vocational or sub-degree education, (6=) lower tertiary education, and (7=) higher tertiary education. Following the supplementary material of \citep{VanNoord2025}, we also created a collapsed education variable, \texttt{education\_3cat}, coded as (1=) less educated for \texttt{eisced} categories 1--2, (2=) middle educated for categories 3--5, and (3=) higher educated for categories 6--7.

Household income is recorded using \texttt{hinctnta}, the ESS household-income decile variable, coded from (1) to (10), where (1) is the lowest income decile and (10) is the highest income decile. Religious belonging is recorded using \texttt{rlgblg}, coded as (1=) yes and (2=) no. Ethnic minority status is recorded using \texttt{blgetmg}, coded as (1=) yes and (2=) no.

Urbanization was derived from the ESS domicile variable. The raw ESS variable \texttt{domicil} is coded from (1=) big city to (5=) farm or home in the countryside. To match the interpretation in the supplementary material, where higher values indicate more urbanised surroundings, we reversed this coding and created \texttt{urbanisation}, coded as (1=) farm or home in the countryside, (2=) country village, (3=) town or small city, (4=) suburbs or outskirts of a big city, and (5=) big city.

Finally, we retained a simple voting-participation variable, \texttt{vote\_simple}, based on the ESS variable \texttt{vote}. This variable is coded as (1=) voted in the last national election, (2=) did not vote, and (3=) not eligible to vote. This variable should not be confused with the more detailed voting-behaviour variable used by \citep{VanNoord2025}, which classifies party choice into loyalty, voice, and exit voting categories using party-vote information and the PopuList classification \citep{Rooduijn2024PopuList}. The detailed voting-behaviour variable was not reconstructed in the present preprocessing step.

\section{Partial-correlation network inference via graphical lasso}
\label{app:lasso}

Let
$V
=
\{1,2,\ldots,p\}$
be the set of all belief nodes and, for a given pair of belief nodes \(i\) and \(j\), the remaining belief nodes are
$C_{ij}
=
V\setminus\{i,j\}$.
The partial-correlation network asks whether belief nodes \(i\) and \(j\) remain associated after the variables in \(C_{ij}\) have been taken into account. In other words, an edge between \(i\) and \(j\) in the partial-correlation network represents a conditional association, not only a simple pairwise association. This is important because two beliefs may appear correlated simply because both are associated with a third belief. The partial-correlation approach reduces this type of indirect association by conditioning on the remaining belief variables.

In the implementation, the same 20 belief variables were selected for the partial-correlation analysis. Missing values were imputed using the median of each belief variable, and the variables were then standardised to have comparable scale. Let \(\widehat{\boldsymbol{\Sigma}}^{(g)}\) denote the estimated covariance matrix of the standardised belief variables in group \(g\). The precision matrix is the inverse covariance matrix:
\begin{equation}
\widehat{\boldsymbol{\Theta}}^{(g)}
=
\left(
\widehat{\boldsymbol{\Sigma}}^{(g)}
\right)^{-1}.
\label{eq:precision_matrix_definition}
\end{equation}
In a Gaussian graphical model, the precision matrix is directly related to conditional associations between variables. If the off-diagonal entry \(\widehat{\Theta}^{(g)}_{ij}\) is zero, then belief nodes \(i\) and \(j\) are conditionally independent given the other belief nodes under the model. If \(\widehat{\Theta}^{(g)}_{ij}\) is nonzero, then the two belief nodes retain a conditional association after accounting for the remaining belief nodes.

The graphical-lasso model estimates a sparse precision matrix, denoted by \(\widehat{\boldsymbol{\Theta}}^{(g)}\). Sparsity is controlled by a regularisation parameter selected by cross-validation in \texttt{GraphicalLassoCV}. The estimated partial correlation between belief nodes \(i\) and \(j\) was computed from the estimated precision matrix as
\begin{equation}
\widehat{P}^{(g)}_{ij}
=
-
\frac{
\widehat{\Theta}^{(g)}_{ij}
}{
\sqrt{
\widehat{\Theta}^{(g)}_{ii}
\widehat{\Theta}^{(g)}_{jj}
}
},
\qquad
i\neq j.
\label{eq:partial_correlation_from_precision}
\end{equation}
Thus, \(\widehat{P}^{(g)}_{ij}\) is the estimated conditional association between belief nodes \(i\) and \(j\), after accounting for the other belief nodes in \(C_{ij}\). The diagonal entries were set to zero,
\begin{equation}
\widehat{P}^{(g)}_{ii}
=
0,
\label{eq:partial_correlation_diagonal}
\end{equation}
because self-associations are not interpreted as network edges.

\section{Comparing belief network inference methods}
\label{app:compare-inference}

We infer the belief networks of these populations using the following approaches: (i) Kendall correlations~\citep{Kendall1938,Masuda2025}, (ii) partial correlations~\citep{EpskampFried2018,Friedman2008}, and (iii) nonparametric Bayesian network reconstruction~\citep{peixotoNetworkReconstructionMinimum2025} (see Methods section).

\begin{figure}
    \centering

    \begin{minipage}[t]{\linewidth}
        \centering
        \includegraphics[width=\linewidth]{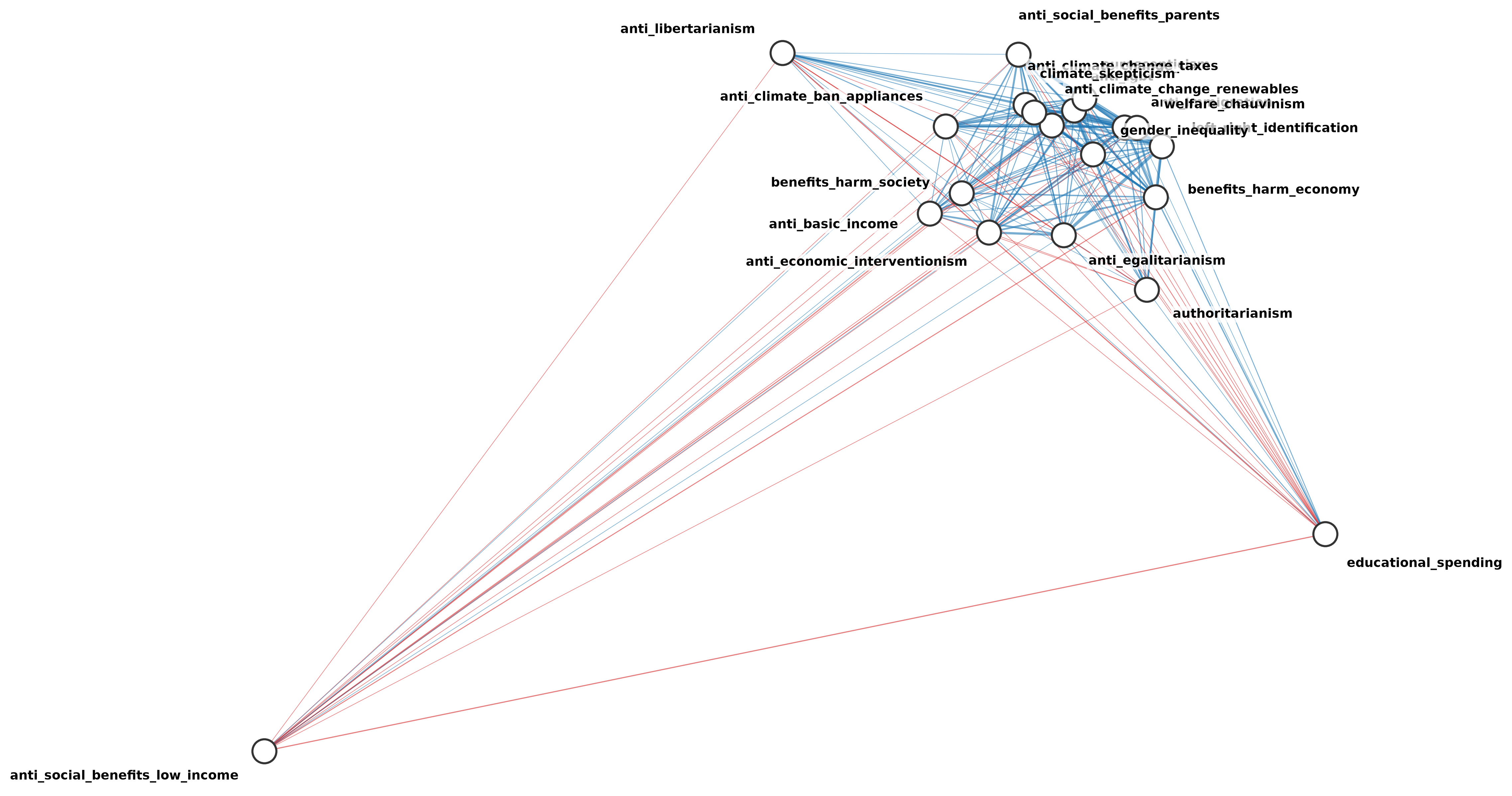}

        \vspace{2mm}
        \small (a)
    \end{minipage}
    \vspace{5mm}
    \hfill
    \begin{minipage}[t]{\linewidth}
        \centering
        \includegraphics[width=\linewidth]{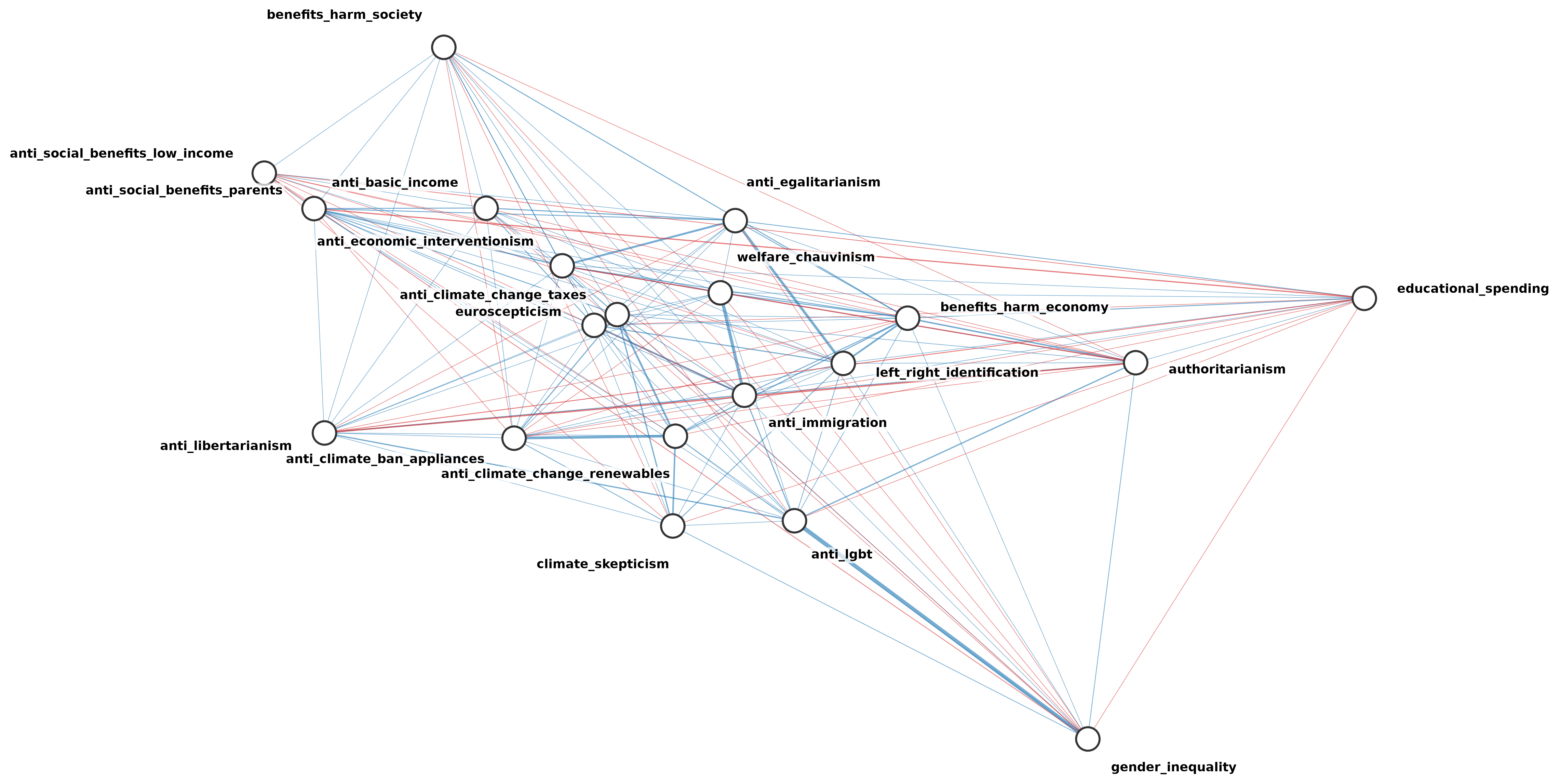}

        \vspace{2mm}
        \small (b)
    \end{minipage}
    \vspace{5mm}

    \caption{Comparison between the belief networks of Great Britain obtained with the Kendall correlation (a) and graphical lasso partial correlation (b) methods. Positive associations are plotted in blue and negative associations in red, with edge thickness proportional to \(\left|W^{(g)}_{jk}\right|\). The network layout is generated using a force-directed spring layout. This layout affects only the visualisation of the network and does not change the inferred edge weights.}
    \label{fig:gb_vs_ru2}
\end{figure}

Correlations and partial correlations are primarily used to understand the overall relational structure of beliefs in our data (Figure~\ref{fig:gb_vs_ru2}). However, it is not straightforward to recover the latent belief network of a population from this relational structure. Networks generated from these methods should be interpreted cautiously. Correlation-based network construction can be sensitive to the choice of association measure, thresholding rule, sample size, and missing-data treatment; therefore, the inferred networks should be interpreted as descriptive belief-association structures rather than as definitive causal maps \citep{Masuda2025,WaldorpMarsman2021}.

\begin{figure}
	\centering
    \includegraphics[width=\linewidth]{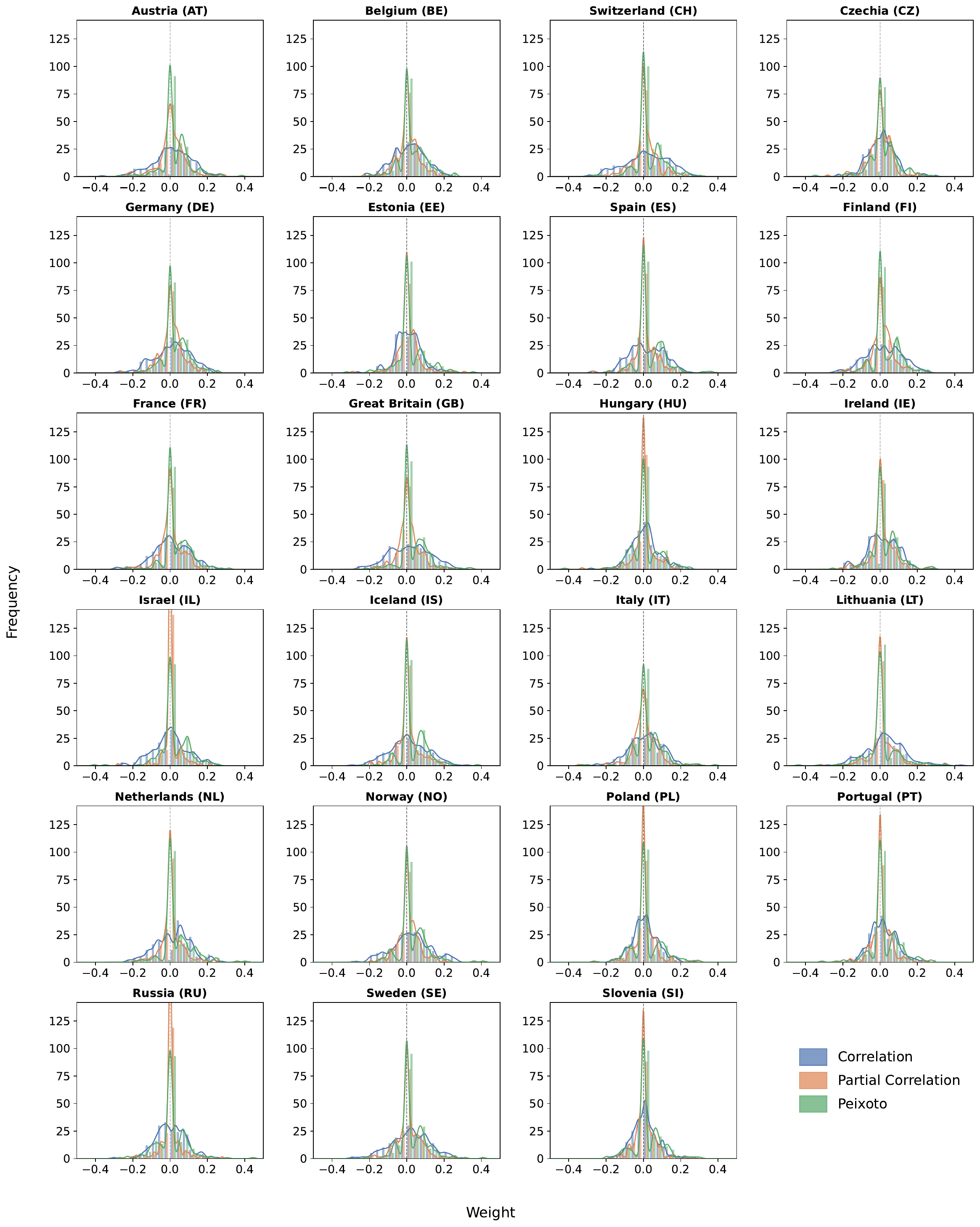}
	\caption{Histogram and a kernel density estimate of weights between beliefs for the three methods that we use (Kendall correlations, partial correlations with graphical lasso and partial correlations using nonparametric Bayesian inference~\citep{peixotoNetworkReconstructionMinimum2025})all 23 countries in our data with the vertical axis representing the frequency of correlation values falling in a bin and the horizontal axis representing the correlations themselves. The densities of the corresponding correlation weight values for each country are represented as well by the curves. A common pattern among all of these distributions is that most correlations are positive and fairly small with only a few strong correlations for graphical lasso and nonparametric Bayesian inference - ideal for inferring belief networks but on the other hand Kendall correlations have a much wider distribution.}
	\label{app:agg_corr_dist}
\end{figure}

\begin{figure}
	\centering
    \includegraphics[width=0.7\linewidth]{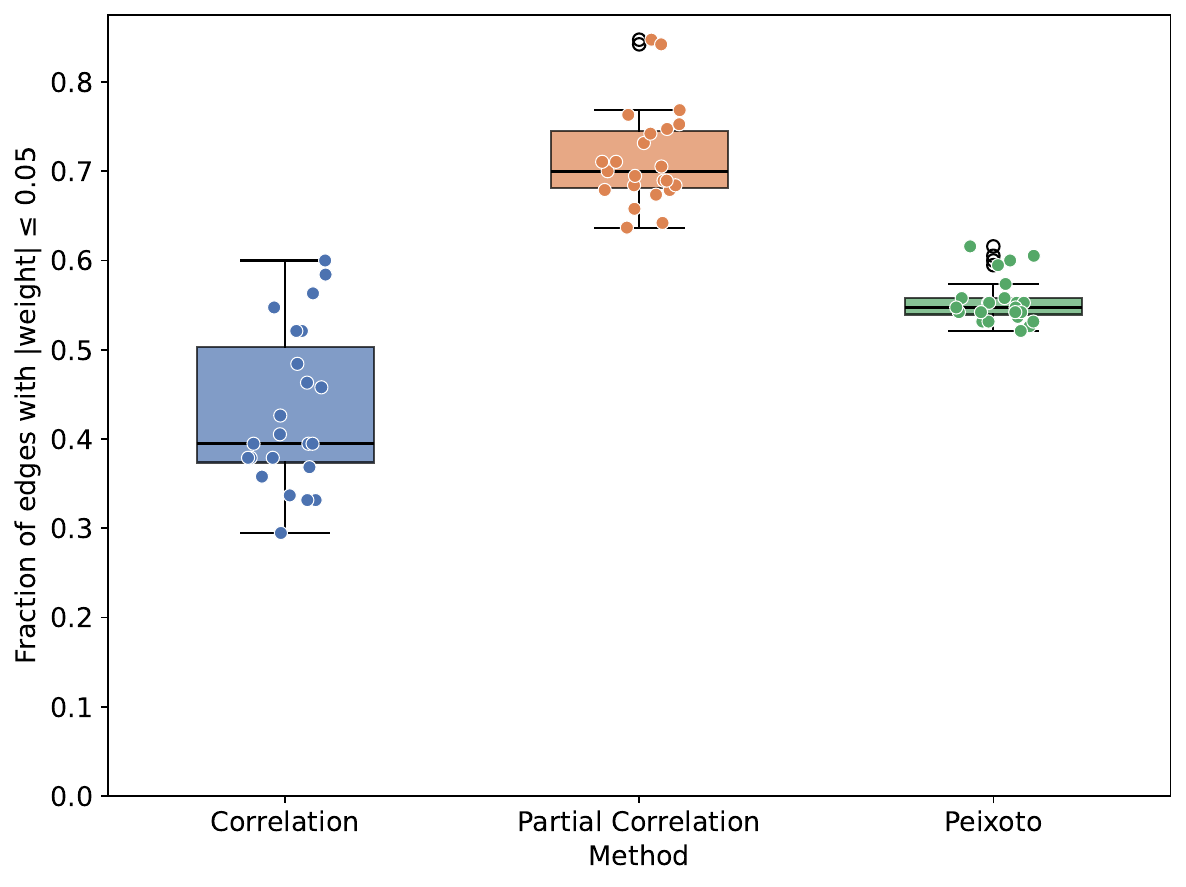}
	\caption{Boxplot of the fraction of near-zero correlation weight values obtained from the three network inference methods we use - Kendall correlations, partial correlations using graphical lasso, partial correlations using nonparametric Bayesian inference~\citep{peixotoNetworkReconstructionMinimum2025}, for all 23 countries in our dataset. Graphical lasso and nonparametric Bayesian inference do a better job of identifying the most important correlations in our data compared to Kendall correlations, as evident from their consistently higher fraction of non-zero values. However, Bayesian inference does a better job of preserving low weight but significant correlations in the data unlike graphical lasso which tends to drive low weights to zero. Moreover, Bayesian inference shows a lower variance of near-zero values across countries and as a result produces more consistent network densities across countries.}
	\label{app:near_zero_boxplot}
\end{figure}

As a result, we use the more principled approach of nonparametric Bayesian network reconstruction~\citep{peixotoNetworkReconstructionMinimum2025} to infer the belief networks of country-wise populations in Europe. This method infers sparse networks without rejecting low weight edges harshly (Figure~\ref{app:agg_corr_dist}, ~\ref{app:near_zero_boxplot}), while also avoiding overfitting. Belief networks inferred using this probabilistic approach serve as the basis for most analyses and simulations in this paper.

\section{The evolution of single beliefs} \label{app:belief_var}
We investigate polarisation at the level of individual beliefs and how they change throughout time. Figures~\ref{fig:UK_all} and~\ref{fig:RU_all} show results for the United Kingdom and Russia tracking the mean and variances of each belief over time. The results where obtained using the minimum description length approach to generate the internal belief network. For the United Kingdom we clearly see the tendencies towards polarisation in the $\beta_{pers}=2$ and $\beta_{soc}=0.5$ regime for nearly every individual belief with the exception of ``educational spending'' and ``anti social benefits low income''. These beliefs correspond with nodes that have more negative correlation in the network then most others (see figures~\ref{fig:gb_vs_ru} and~\ref{fig:gb_vs_ru2}). Polarisation is much less prominent in the results for Russia, where beliefs also tend to grow more extreme but not necessarily in opposing directions.
\begin{figure}      
  \centering
  \includegraphics[width=0.8\linewidth]{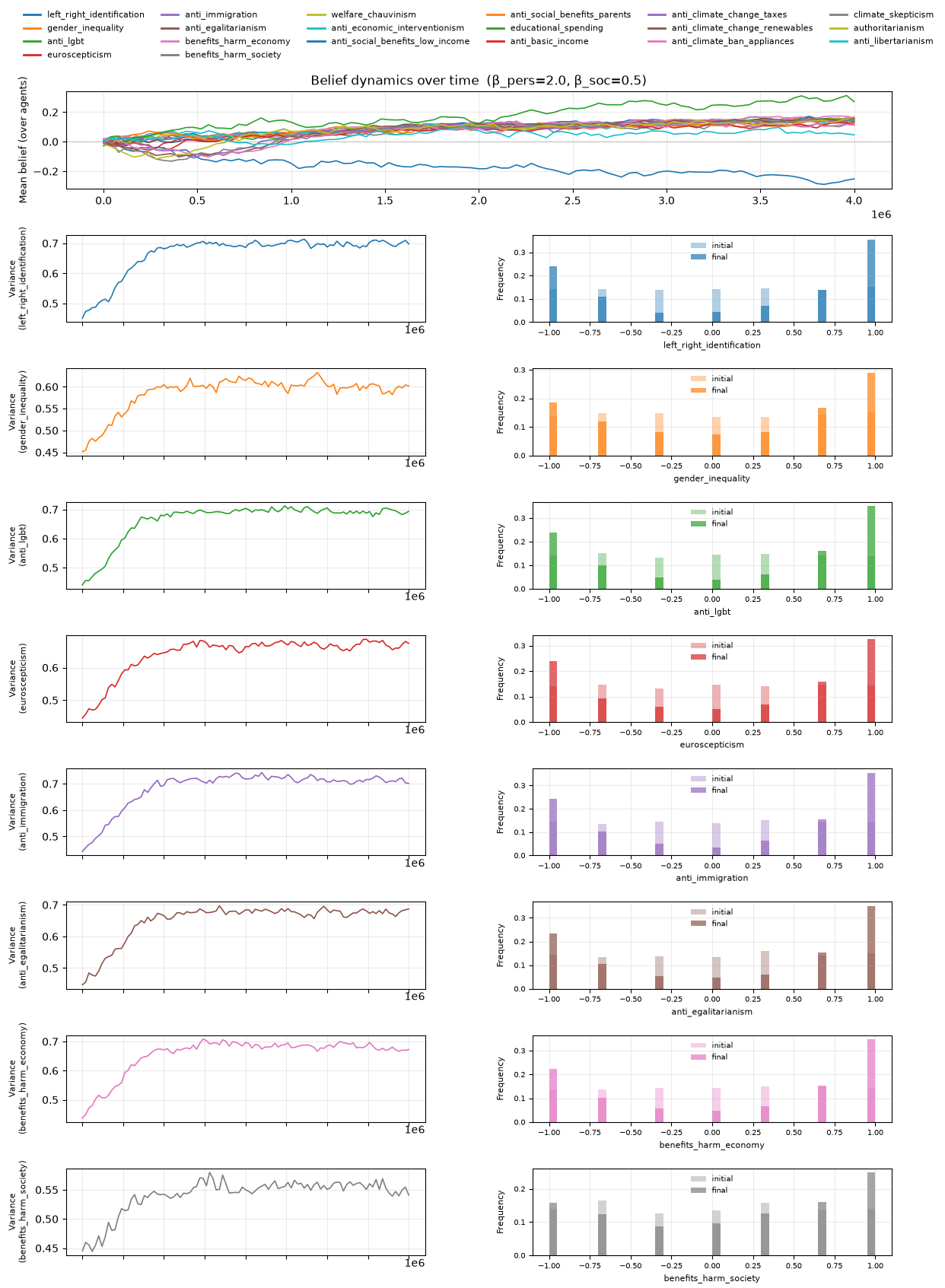}
  \caption{Variances over time (left column) and the distribution at the final state (right column) for all belief after an ABM run for the United Kingdom with parameters $\beta_{pers}=2$ and $\beta_{soc}=0.5$.} 
  \label{fig:UK_all}
\end{figure}

\begin{figure}
  \ContinuedFloat
  \centering
  \includegraphics[width=0.8\linewidth]{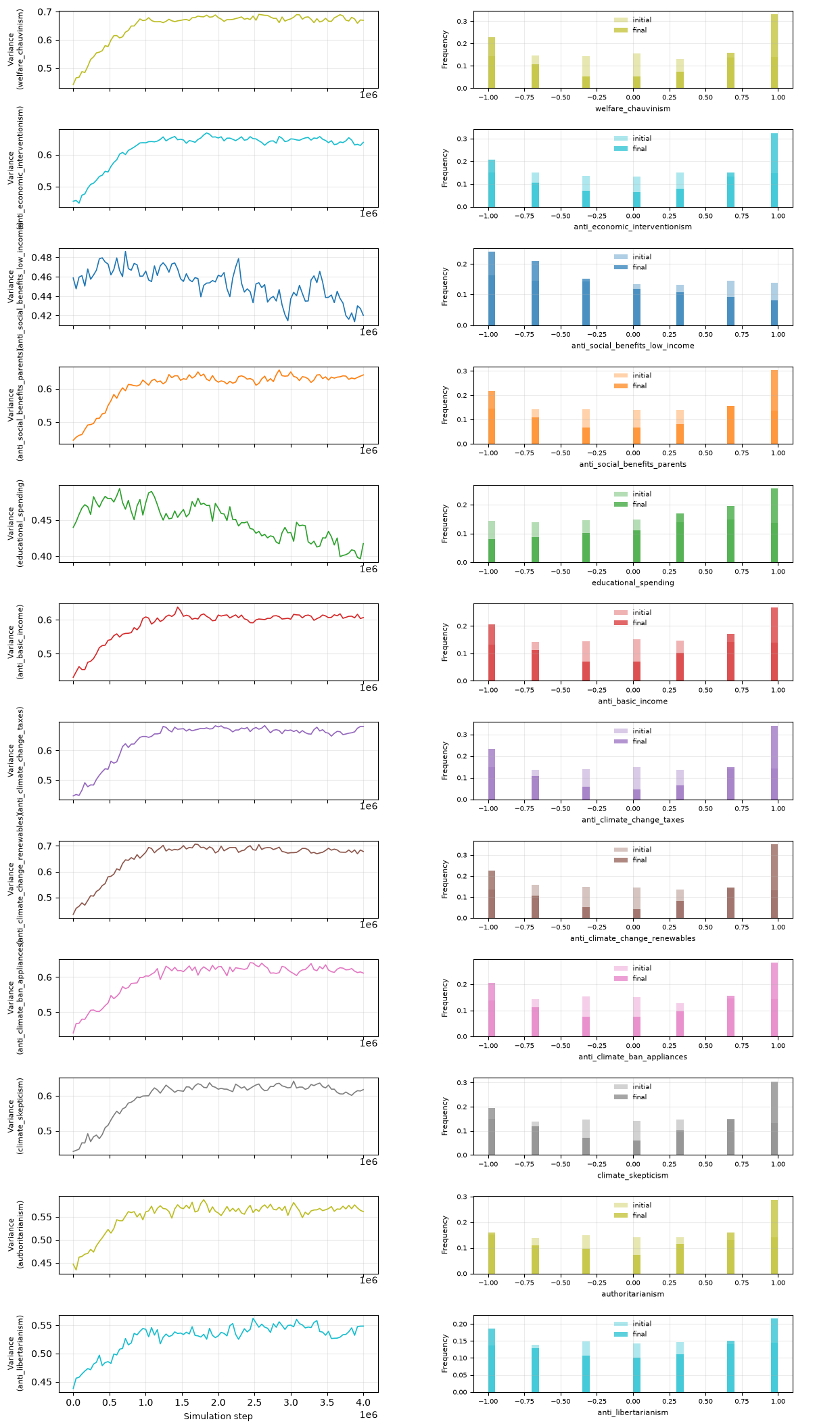}
  \caption{(continued)}
\end{figure}

\begin{figure}      
  \centering
  \includegraphics[width=0.8\linewidth]{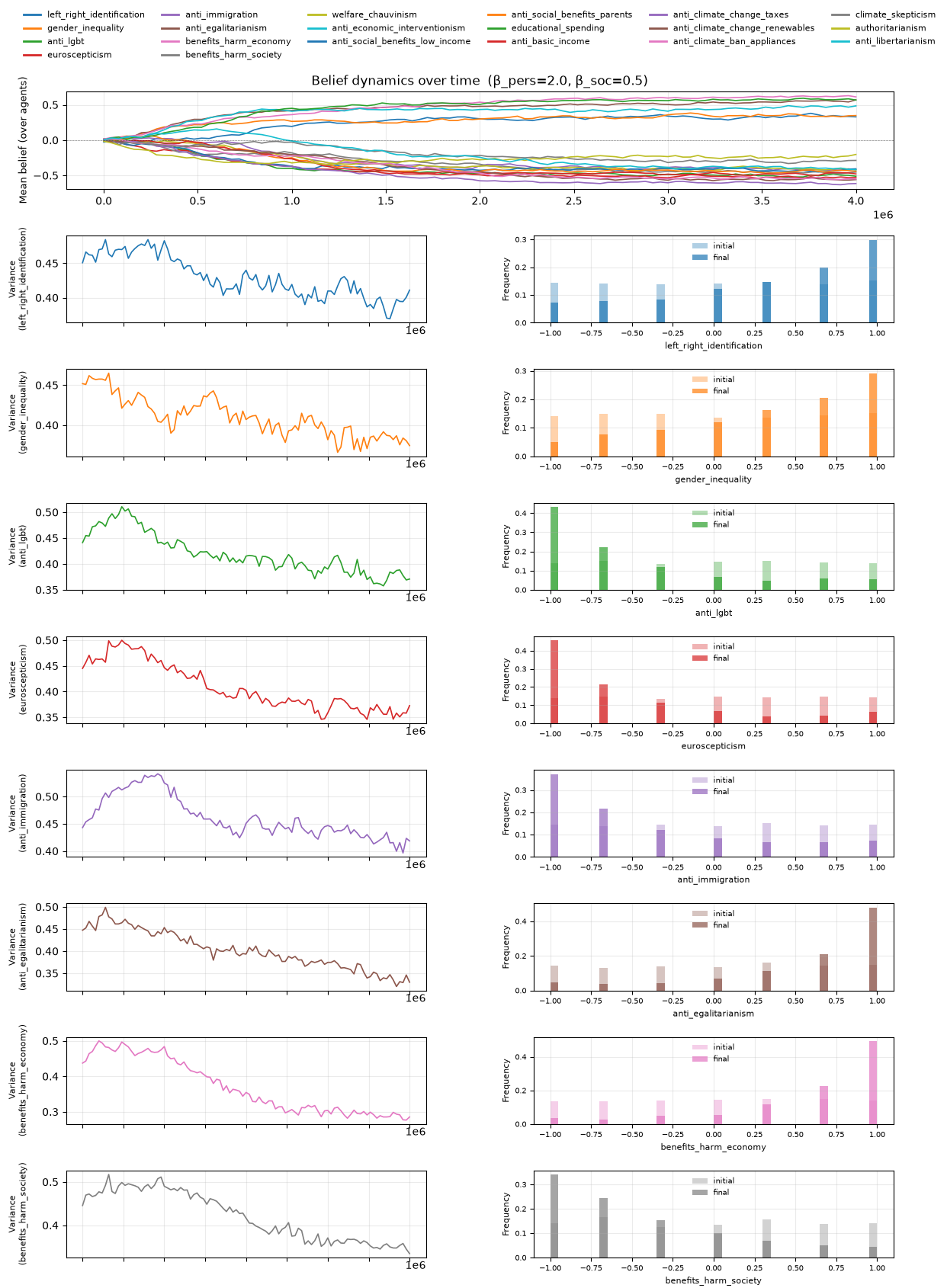}
  \caption{Variances over time (left column) and the distribution at the final state (right column) for all belief after an ABM run for Russia with parameters $\beta_{pers}=2$ and $\beta_{soc}=0.5$.}
  \label{fig:RU_all}
\end{figure}

\begin{figure}
  \ContinuedFloat
  \centering
  \includegraphics[width=0.8\linewidth]{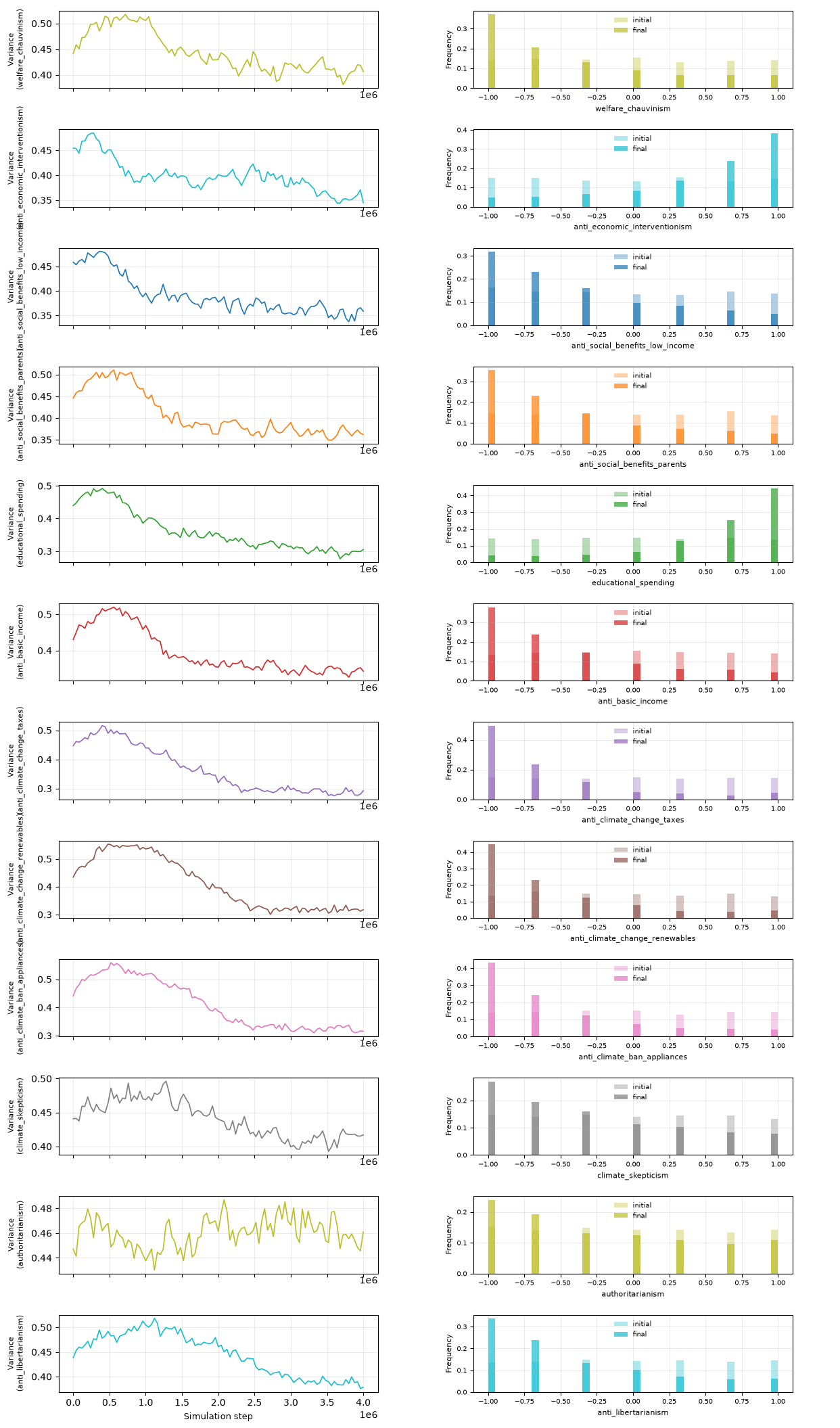}
  \caption{(continued)}
\end{figure}
\FloatBarrier

\newpage
\section{Extended analysis of polarisation measures}
\label{app:polarisation-measures}

This appendix provides additional details on the polarisation measures discussed in the Methods section. We first summarise one-dimensional distributional measures, including the (m)-value, and then describe how these ideas can be extended to multidimensional opinion vectors. Finally, we report the effective-rank analysis used as an additional robustness check for the inferred opinion-dependence matrices.

\subsection{One-dimensional polarisation measures}
\label{app:onedim_polarisation}

In the one-dimensional case, polarisation is not a single unambiguous property. Ref.~\citep{bramson2017understanding} distinguishes nine different conceptual ways of measuring polarisation in one-dimensional opinion distributions:

\begin{enumerate}
    \item Spread --- the distance between the most extreme opinions
    
    \item Dispersion or statistical variation --- the statistical spread of opinions, for example standard deviation or mean absolute deviation.

    \item Coverage --- the fraction of occupied intervals on the opinion scale; lower coverage indicates that opinions are concentrated in narrower regions.

    \item Regionalisation --- the number of empty gaps between occupied regions of the distribution.

    \item Community fracturing --- the number of groups or clusters in the distribution.

    \item Distinctness --- the degree to which groups are separated from one another.

    \item Group divergence --- the distance between group means or other central characteristics of groups.

    \item Group consensus or solidarity --- the degree of agreement within each group; lower within-group variation corresponds to higher polarisation.

    \item Size parity --- the balance between group sizes; polarisation is higher when groups are comparable in size.
\end{enumerate}

These nine senses of polarisation are not mutually implied. For example, a distribution may exhibit strong within-group consensus but have a small overall spread, or it may have large divergence between group means while the groups themselves are highly unequal in size. At the same time, this does not mean that all measures are fully independent in every possible setting: higher-order relations between particular combinations of measures may exist once additional properties of the distribution are fixed. A systematic analysis of such higher-order dependencies is beyond the scope of the present work. Thus, one-dimensional polarisation measures should be interpreted as complementary diagnostics rather than interchangeable definitions of the same quantity.

For histograms constructed from one-dimensional opinion distributions, another commonly used diagnostic is the (m)-value. This measure was proposed by Ref.~\citep{gregg2019frequency} and was also used in Ref.~\citep{prasetya2020model}. It captures the modality of a distribution and is defined as
\begin{equation}
m\text{-value}(x)
=
\frac{1}{M}
\sum_{i=2}^{n}
\left|x_i-x_{i-1}\right|,
\end{equation}
where (x) is a histogram with (n) bins and (M) is its maximum bin height. The more pronounced the peaks and valleys of the histogram, the larger the (m)-value. A perfectly unimodal distribution yields (m\text{-value}=2), whereas a perfectly bimodal distribution with two peaks of the same height yields (m\text{-value}=4). Values larger than (4) indicate that the distribution is spread across multiple peaks and is therefore not necessarily polarised in a bimodal sense. Following Ref.~\citep{gregg2019frequency}, a value around $2.4$ can be used as an approximate threshold for identifying bimodality.

The (m)-value is most appropriate for relatively simple low-dimensional settings, especially when polarisation is expected to appear as unimodality or bimodality in a single opinion distribution. This is not the main situation considered in the present study. Here, each agent is represented by a vector of interdependent beliefs, and the relevant structure is not only the marginal distribution of each belief, but also the joint configuration of beliefs across the full belief space. Therefore, applying the (m)-value separately to individual belief dimensions would provide only a partial diagnostic and would not capture the multidimensional structure of polarisation generated by the model.

\subsection{Multidimensional opinion vectors and belief-space distance}
\label{app:multidim-distance}

A natural generalisation of the one-dimensional case is to represent each agent by a multidimensional opinion vector. Such representations are well established in the literature on opinion dynamics, especially in models where agents hold positions on several interdependent topics. What is less standardised, however, is the measurement of polarisation as a single scalar quantity in a genuinely multidimensional opinion space.

One example of a genuinely multidimensional polarisation measure is the approach described in Ref.~\citep{gubanov2019multidimensional}. In that model, each agent's opinion is represented as a stochastic vector over several extreme alternatives, and polarisation is measured through a multidimensional index based on the average distance of agents' opinions from the population centre of mass. This approach captures polarisation directly in a multidimensional opinion space. However, it requires the explicit specification of several extreme alternatives, or effectively predefined opinion clusters. More generally, this class of approaches requires either an explicit labelling of opinion groups or a separate clustering procedure that identifies such groups from the data. Since our model does not assume predefined alternatives or externally labelled opinion groups, and since introducing a clustering step would add an additional modelling layer, we do not use this type of multidimensional polarisation index explicitly in the present work.

Multidimensional opinion vectors are also used in other models of opinion dynamics. For example, \cite{parsegov2016novel} represent each agent's opinion as a vector of topic-specific components and introduce the MiDS matrix (C), which encodes interdependencies between issues. Positive couplings in this matrix tend to align topic-specific opinions, whereas negative couplings may lead to their polarisation. Similarly, \cite{baumann2021emergence} consider opinions as vectors in a multidimensional topic space and analyse how correlations between different opinion dimensions may emerge.

However, in such works, a general scalar measure of multidimensional polarisation that is essentially different from one-dimensional distributional measures is usually not introduced. Instead, the analysis is typically based on a collection of one-dimensional diagnostics applied to different coordinates of the opinion vector, together with pairwise relations between these coordinates, such as correlations or topic-overlap parameters.

In the simulations studied in this paper, each agent is represented by a vector of (m) beliefs,
\begin{equation}
\mathbf{b}^{k}
=
\left(
b_1^k,\ldots,b_m^k
\right),
\end{equation}
where (k) indexes agents and (i=1,\ldots,m) indexes belief dimensions. A direct way to measure disagreement between two agents in the full belief space, without imposing predefined groups or clusters, is to compute the Euclidean distance between their belief vectors. For every pair of agents ((k,l)), we define
\begin{equation}
d_{kl}
=
\left|
\mathbf{b}^{k}-\mathbf{b}^{l}
\right|*2
=
\sqrt{
\sum*{i=1}^{m}
\left(
b_i^k-b_i^l
\right)^2
}.
\label{eq:app_pairwise_belief_distance}
\end{equation}
This distance is small when two agents hold similar positions across all belief dimensions and large when their belief vectors are far apart in the multidimensional belief space. The main polarisation measure used in the simulations is then defined as the variance of these pairwise distances:
\begin{equation}
P_{\mathrm{MD}}
=
\operatorname{Var}
\left(
{d_{kl}: k<l}
\right).
\label{eq:app_multidim_polarisation}
\end{equation}
This quantity is a heuristic measure of polarisation across the full belief space. It does not require the prior specification of opinion groups, extreme alternatives, or ideological poles. Instead, it captures how heterogeneous the population becomes in terms of the distances between agents' complete belief vectors.

This measure differs from applying one-dimensional diagnostics separately to each belief coordinate. Component-wise diagnostics can show whether individual beliefs become more dispersed or more bimodal, but they do not directly capture how beliefs combine across dimensions. The distance-based measure instead treats the full belief vector as the object of comparison.

\subsection{Effective rank of inferred opinion-dependence matrices}
\label{app:effective-rank}

In addition to the analysis of distances between opinion vectors, one can also study the global structure of opinion-dependence matrices. Given an opinion matrix
\begin{equation}
X \in \mathbb{R}^{n \times m},
\end{equation}
where rows correspond to agents and columns correspond to issues or beliefs, a correlation or dependence matrix between opinion dimensions can be constructed and its structural properties analysed. In this setting, polarisation can be described indirectly in terms of dimensional compression. If many issue or belief positions become aligned with a small number of latent ideological axes, the opinion space effectively collapses to a lower-dimensional structure. This idea is close to the notion of issue alignment discussed by Ref.~\citep{schweighofer2024raising}, where polarisation is associated with the alignment of issue positions into a common ideological spectrum.

A related approach is used by Ref.~\citep{vendeville2026political}, who analyse the latent dimensionality of political attitudes and show that stronger ideological polarisation can be associated with a lower-dimensional attitude structure. In this sense, latent dimensionality can be interpreted as the number of effective directions along which political attitudes vary: lower dimensionality indicates that many issues are aligned with a smaller number of ideological axes.

In our setting, each country is associated with its own inferred opinion-dependence matrix. Therefore, in order to summarise the results across countries, it is necessary to divide countries into two groups. As in the main text, we refer to these groups as conditionally `Western'' and conditionally `Eastern'' Europe, while keeping in mind that this division is not meant to be strictly geographical and may include outliers or semantic drift in the position of some countries.

For reproducibility, we used the following procedure to obtain the two groups. First, we applied K-means++ clustering~\citep{arthur2007k} with (K=2) to the country-level opinion vectors. Then, for the resulting clustering, we computed the silhouette score~\citep{rousseeuw1987silhouettes}. For an observation (i), let (a(i)) be the average distance from (i) to all other observations in the same cluster, and let (b(i)) be the smallest average distance from (i) to observations in any other cluster. The silhouette coefficient of observation (i) is defined as
\begin{equation}
s(i)
=
\frac{b(i)-a(i)}
{\max{a(i),b(i)}}.
\end{equation}
The average silhouette score is then
\begin{equation}
S
=
\frac{1}{N}
\sum_{i=1}^{N} s(i),
\end{equation}
where (N) is the number of observations. Values of (S) close to (1) indicate well-separated clusters, values close to (0) indicate overlapping clusters, and negative values indicate that some observations may be assigned to an inappropriate cluster.

\begin{figure}
\centering
\includegraphics[width=0.95\linewidth]{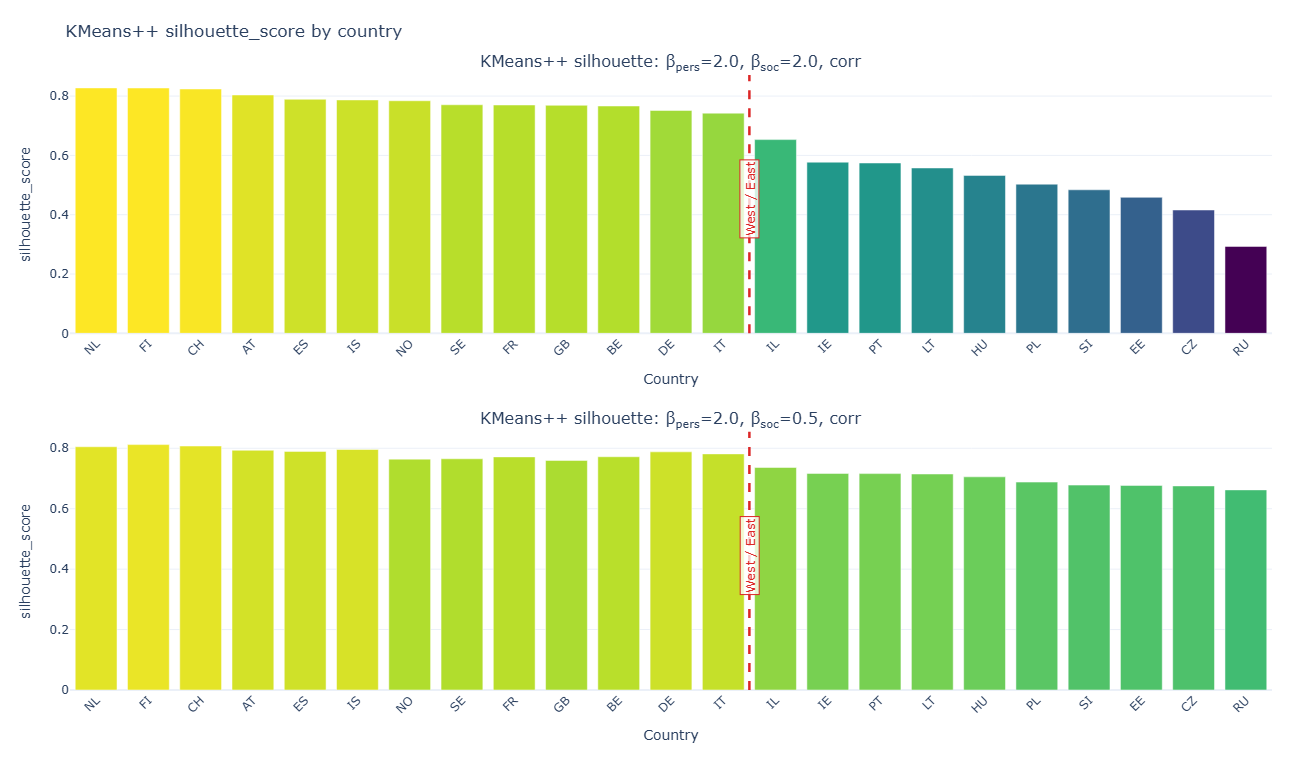}
\caption{K-means++ clustering of country-level opinion vectors and the corresponding silhouette-based separation. The figure illustrates the two-group split used in the effective-rank analysis.}
\label{fig:appendix_silhouette}
\end{figure}

Independently of the specific clustering method and of the particular way in which the opinion-dependence network is inferred (correlation matrices, partial-correlation matrices, or the reconstruction approach based on minimum description-length of~\citep{peixotoNetworkReconstructionMinimum2025}) the main analysis suggests that the inferred opinion structures should differ systematically between the two groups of countries. 
In particular, we expect the opinion-dependence matrices of conditionally ``Western'' and conditionally ``Eastern'' countries to exhibit different latent dimensional structures. 
For illustration, the clustering obtained from correlation-based networks shows the expected separation: for sufficiently large $\beta_{\mathrm{pers}}$ (in our experiment, $\beta_{\mathrm{pers}}=2.0$) and varying values of $\beta_{\mathrm{soc}}$, the countries are visually divided into two groups of comparable size, with the conditionally ``Eastern'' cluster containing Israel, Ireland, Portugal, Lithuania, Hungary, Poland, Slovenia, Estonia, Czechia, and Russia.

Using this division, we computed the effective ranks of the inferred opinion-dependence matrices obtained by all three methods described in the Methods section. In many applications, latent dimensionality is estimated from the eigenvalue spectrum of a positive semidefinite covariance or correlation matrix, as in PCA. In that case, eigenvalues are nonnegative and can be interpreted as variance contributions along orthogonal directions. However, in our case the inferred matrices are not necessarily positive semidefinite. This is especially important for partial-correlation matrices and for the matrices reconstructed by the approach of~\citet{peixotoNetworkReconstructionMinimum2025}, which in our experiments were frequently not positive definite. Consequently, a standard PCA-like interpretation of the positive eigenvalue spectrum would be inappropriate: negative eigenvalues may occur, and the eigenvalues cannot be interpreted directly as explained variances. Therefore, instead of using PCA-based dimensionality, we compute the effective rank of the matrices in the sense of~\citet{roy2007effective}, using the absolute eigenvalue spectrum.

Since the inferred matrices considered here are correlation-type matrices, they are symmetric. Thus, if
\begin{equation}
A = Q \Lambda Q^\top
\end{equation}
is the eigendecomposition of such a matrix, its singular values are equal to the absolute values of its eigenvalues:
\begin{equation}
\sigma_i(A)=|\lambda_i(A)|.
\end{equation}
For this reason, we explicitly used the moduli of eigenvalues. Let
\begin{equation}
|\lambda_1|,\ldots,|\lambda_q|
\end{equation}
be the nonzero absolute eigenvalues of (A), where ($q=\operatorname{rank}(A)$). We define the normalised spectral weights as
\begin{equation}
p_i
=
\frac{|\lambda_i|}
{\sum_{j=1}^{q}|\lambda_j|}.
\end{equation}
The effective rank is then
\begin{equation}
\operatorname{erank}(A)
=
\exp\left(
-\sum_{i=1}^{q}
p_i \log p_i
\right).
\end{equation}
This quantity is close to one when the matrix is effectively dominated by a single latent direction and increases when several independent directions contribute comparably. For positive semidefinite matrices, the same expression coincides with the usual entropy-based effective dimensionality computed from the nonnegative eigenvalue spectrum.

Previous work suggests that stronger ideological polarisation is often associated with a lower latent dimensionality of attitudes: issue positions become aligned along a smaller number of ideological dimensions. For example, \citet{schweighofer2024raising} argue that polarisation requires the alignment of issue positions into a common ideological spectrum, while \citet{vendeville2026political} show, in a setting related to the left--right divide, that the dimensionality of attitude distributions shrinks as ideological polarisation increases.

\begin{figure}
\centering
\includegraphics[width=0.95\linewidth]{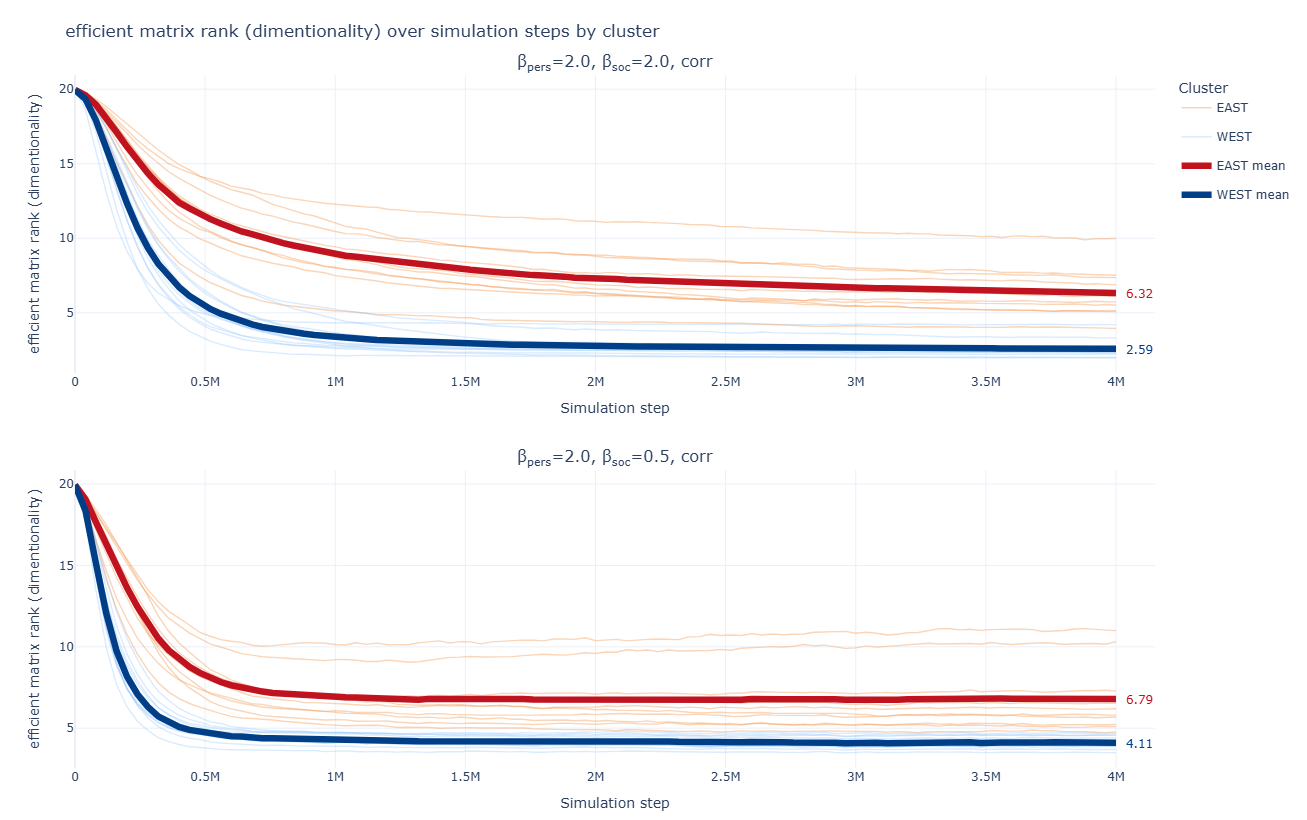}
\caption{Effective ranks of correlation-based opinion-dependence matrices for the two country groups. Lower effective rank indicates a more compressed latent opinion structure.}
\label{fig:appendix_erank_cor}
\end{figure}

We first computed the effective ranks for the correlation-based opinion-dependence matrices. This case is the closest to the standard interpretation of latent dimensionality, since the matrices directly describe pairwise dependencies between opinion dimensions. The resulting values provide a baseline against which the more restrictive partial-correlation and MDL-based reconstructions can be compared. As shown in Figure~\ref{fig:appendix_erank_cor}, the two country groups differ in the effective dimensionality of their correlation structures: lower values indicate that the corresponding opinion matrices are dominated by fewer latent directions.

\begin{figure}
\centering
\includegraphics[width=0.95\linewidth]{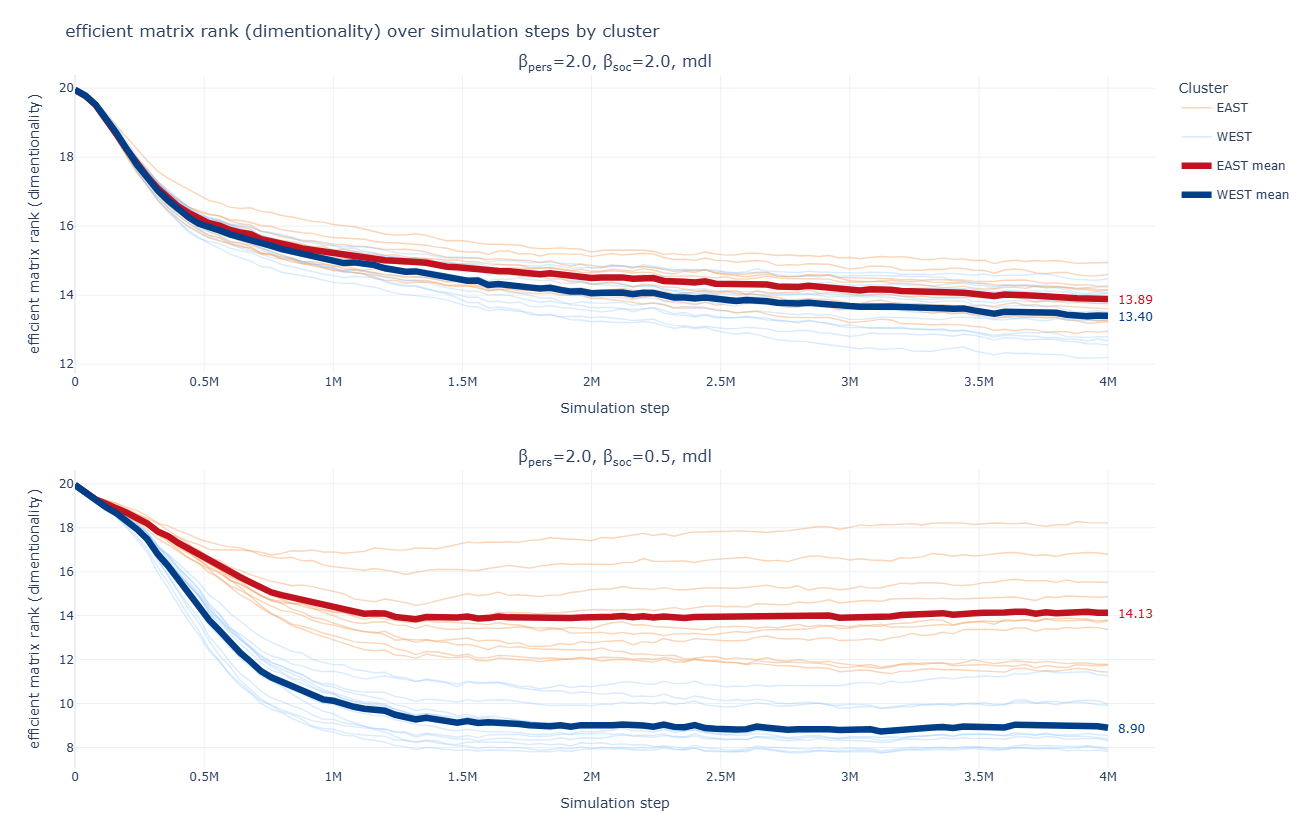}
\caption{Effective ranks of opinion-dependence matrices inferred using the minimum-description-length-based reconstruction approach.}
\label{fig:appendix_erank_mdl}
\end{figure}

The same analysis was repeated for matrices inferred using the minimum-description-length-based network reconstruction approach. This construction is more selective and retains only dependencies supported by the reconstruction procedure. Nevertheless, the qualitative separation between the two groups remains visible, as shown in Figure~\ref{fig:appendix_erank_mdl}.

\begin{figure}
\centering
\includegraphics[width=0.95\linewidth]{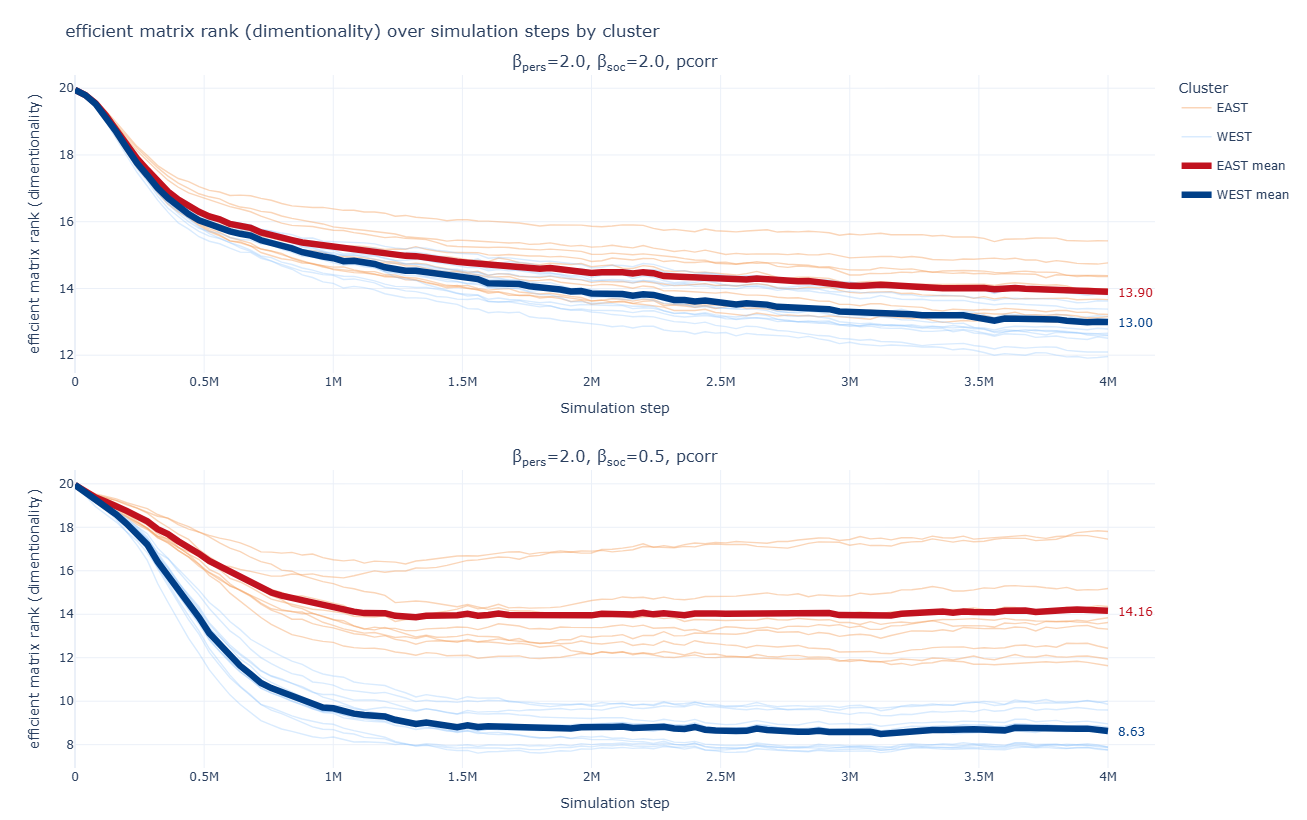}
\caption{Effective ranks of partial-correlation opinion-dependence matrices for the two country groups.}
\label{fig:appendix_erank_pcor}
\end{figure}

Finally, we computed effective ranks for partial-correlation matrices. Partial correlations remove indirect pairwise associations mediated by other opinion dimensions and therefore provide a more conservative estimate of direct dependencies between opinions. The resulting effective ranks, shown in Figure~\ref{fig:appendix_erank_pcor}, again support the same qualitative picture.

\FloatBarrier

\end{document}